\documentclass[aps,twocolumn,prb,amsmath,amssymb]{revtex4-2}

\usepackage{amsmath}
\usepackage{amssymb}
\usepackage[varg]{txfonts}
\usepackage{color}
\usepackage[unicode=true,
bookmarks=false,
breaklinks=false,pdfborder={0 0 1},backref=false,colorlinks=false]
{hyperref}
\usepackage{tikz}
\usetikzlibrary{shapes}
\usepackage{xcolor}
\hypersetup{colorlinks}

\usepackage{graphicx}
\usepackage{bm}

\makeatletter
\newcommand\bigDiamond{\mathop{\mathpalette\bigDi@mond\relax}}
\newcommand\bigDi@mond[2]{\vcenter{\hbox{\m@th \scalebox{\ifx#1\displaystyle 2\else1.2\fi}{$#1\Diamond$}}}}
\newcommand{\RNum}[1]{\uppercase\expandafter{\romannumeral #1\relax}}

\def\XXint#1#2#3{{\setbox0=\hbox{$#1{#2#3}{\int}$}
		\vcenter{\hbox{$#2#3$}}\kern-.5\wd0}}

\def\be{\begin{equation}}
	\def\ee{\end{equation}}

\def\bi{\begin{itemize}}
	\def\ei{\end{itemize}}
\def\bn{\begin{enumerate}}
	\def\en{\end{enumerate}}
\def\bea{\begin{eqnarray}}
	\def\eea{\end{eqnarray}}
\newcommand{\bpm}{\begin{pmatrix}}
	\newcommand{\epm}{\end{pmatrix}}

\def\ba{\begin{array}}
	\def\ea{\end{array}}
\def\bd{\begin{displaymath}}
	\def\ed{\end{displaymath}}

\renewcommand{\imath}{\hspace{1pt}\mathrm{i}\hspace{1pt}}

\renewcommand{\vec}{\mathbf}
\renewcommand{\Re}{\mathop{\mathrm{Re}}\nolimits}
\renewcommand{\Im}{\mathop{\mathrm{Im}}\nolimits} 

\begin{document}
	
	\title{Probing the collective excitations of excitonic insulators in an optical cavity}
	
	\author{Elahe Davari}
	\affiliation{Department of Physics, Sharif University of Technology, Tehran 14588-89694, Iran}
	\author{Mehdi Kargarian}
	\email{kargarian@sharif.edu}
	\affiliation{Department of Physics, Sharif University of Technology, Tehran 14588-89694, Iran}
	\date{\today}
	
	\begin{abstract}
		The light--matter interaction in optical cavities offers a promising ground to create hybrid states and manipulate material properties. In this work, we examine the effect of light-matter coupling in the excitonic insulator phase using a quasi one-dimensional lattice model with two opposite parity orbitals at each site. We show that the model allows for a coupling between the collective phase mode and cavity photons. Our findings reveal that the collective mode of the excitonic state significantly impacts the dispersion of the cavity mode, giving rise to an avoiding band crossing in the photon dispersion. This phenomenon is absent in trivial and topological insulator phases and also in phonon-mediated excitonic insulators, underscoring the unique characteristics of collective excitations in excitonic insulators. Our results demonstrate the significant impact of light-matter interaction on photon propagation in the presence of excitonic collective excitations. 
	\end{abstract}
	\maketitle
	
	\section{Introduction}
	The condensation of fermionic bound states in macroscopic quantum states and  collective dynamics are among the fascinating phenomena, featuring the complexity of the ground state of correlated systems. A prime example is the excitonic insulator, where excitons - the bound states of electron-hole pairs due to Coulomb interaction - coherently form a condensate, which exhibits superfluid-like behavior with collective Higgs and Goldstone modes \cite{jerome1967excitonic,kohn1967excitonic,keldysh2024collective,halperin1968excitonic,halperin1968possible}. Despite being predicted theoretically over fifty years ago \cite{jerome1967excitonic,keldysh2024collective,kohn1967excitonic}, the material discovery and experimental verification of exciton condensation have been challenging for decades. While the early observations of condensation were reported in a bilayer semiconductor system at very low temperatures \cite{butov2002macroscopically,eisenstein2004bose,wang2019evidence,nandi2012exciton},   certain transition metal chalcogenides have been recently identified as promising candidates with critical transition around the room temperature \cite{sunshine1985structure,wakisaka2009excitonic,kaneko2012excitonic,seki2014excitonic,monney2009spontaneous,monney2012electron,kaneko2018exciton}. 
	
	In spite of extensive works done in recent years, the very nature of the excitonic phase in these materials has not been conclusively identified, and the full understanding of the nature of ground state is still lacking. One famous example is \(\mathrm{Ta_2NiSe_5}\) \cite{sunshine1985structure,wakisaka2009excitonic,kaneko2012excitonic,seki2014excitonic}, which exhibits excitonic condensation and structural phase transitions concurrently when the temperature falls below the transition point. Both mechanisms result in a  gap opening in energy spectrum. Additionally, strong electron-electron and electron-phonon interactions in such compounds raise questions about the nature of the phase transition and gap opening \cite{mor2017ultrafast,okazaki2018photo,tang2020non,baldini2023spontaneous,kim2020phononic,bretscher2021ultrafast}. To unravel the collective properties of the exciton condensation, one approach which has been utilized extensively in recent years is to drive the system out of equilibrium using the laser pulses and probe the excitations. In almost all of these works the optical pulses are in the classical regime and their response is used to infer the correlations underlying the excitations \cite{davari2024optical,mor2017ultrafast,bretscher2021imaging,baldini2023spontaneous,golevz2022unveiling}.
	
	Use of optics in quantum regime, i.e., optical processes involving single or multiple photon modes in quantum cavities, may offer yet another means to generate electron-photon entangled states through the light-matter interactions. These states can potentially show interesting phenomena and reveal intricate properties of materials
	\cite{sentef2020quantum,ruggenthaler2018quantum,ruggenthaler2018quantum,ruggenthaler2018quantum},  suggesting the quantum cavities as a powerful tool for studying the phase space of materials. It provides a deeper understanding of material properties and uncovers physical phenomena not accessible in classical optical-based methods \cite{lysne2023quantum,dmytruk2022controlling,passetti2023cavity,lenk2020collective,ruggenthaler2018quantum,frisk2019ultrastrong,schlawin2022cavity}. 
	
	The chief goal of the current study is to investigate the spectrum of an excitonic insulator when coupled to the light in a cavity. While the previous works mainly focus on the stabilization of condensate in cavities \cite{lenk2020collective,mazza2019superradiant, andolina2019cavity}, here we particularize the study to the collective modes by introducing a model, which allows for coupling between phase and cavity photon modes. We consider a one-dimensional model with two orbitals of opposite parities at each site \cite{shockley1939surface}, the so-called s-p chain, where in the presence of local Coulomb interaction the phase diagram shows three insulating phases: excitonic insulator, topological and trivial insulators \cite{khatibi2020excitonic,kunevs2015excitonic}, and also a phonon-mediated excitonic insulator. Thus, the model provides a fertile ground to explore the interplay between different types of ground states and cavity modes. This is important because as we mentioned above the true ground state of \(\mathrm{Ta_2NiSe_5}\) is still controversial. The setup studied in this work may help envisage responses to be explored in experiments. In particular, we are interested to understand how do the collective phase modes affect photon propagation in the cavity? and keeping an eye on the potential future experiments, we try to see how the photon characteristics inside the cavity can be measured in, e.g., heterodyne photodetector \cite{vogel2006quantum,loudon2000quantum,lysne2023quantum,grynberg2010introduction}. To address these questions, we calculate the photon spectral function within the Random Phase Approximation (RPA) to examine the hybrid light-matter states. The  results indicate that the collective modes of the excitonic insulator lead to a light-matter entangled state and create a gap in the propagation of the cavity photon mode, while in other insulators the photon dispersion essentially remains intact.

	This paper is organized as follows. In Sec.\ref{sec:model and method}, we review the one-dimensional s-p chain model. In Sec.\ref{sec:light-matter}, the formalism of light-matter interaction is presented. The coupling to optical modes of cavity is presented in  Sec.\ref{sec:Spectroscopic Analysis of the Cavity Mode}. In Sec.\ref{sec:hybrid_modes}, the spectroscopy of the cavity mode is studied. In Sec. \ref{sec:phonon}, we discuss the response of the Coulomb- and phonon-mediated pairing in the excitonic insulator. We conclude in Sec.\ref{sec:conclusion}, and the details of the cavity photon Green’s function calculations and the collective excitonic mode spectrum are relegated to the appendices. The data that support the findings of this article are openly available \cite{DataAvailiblity_CavityEI}.
	
	\begin{figure}[t]
		\includegraphics[width=0.4\textwidth]{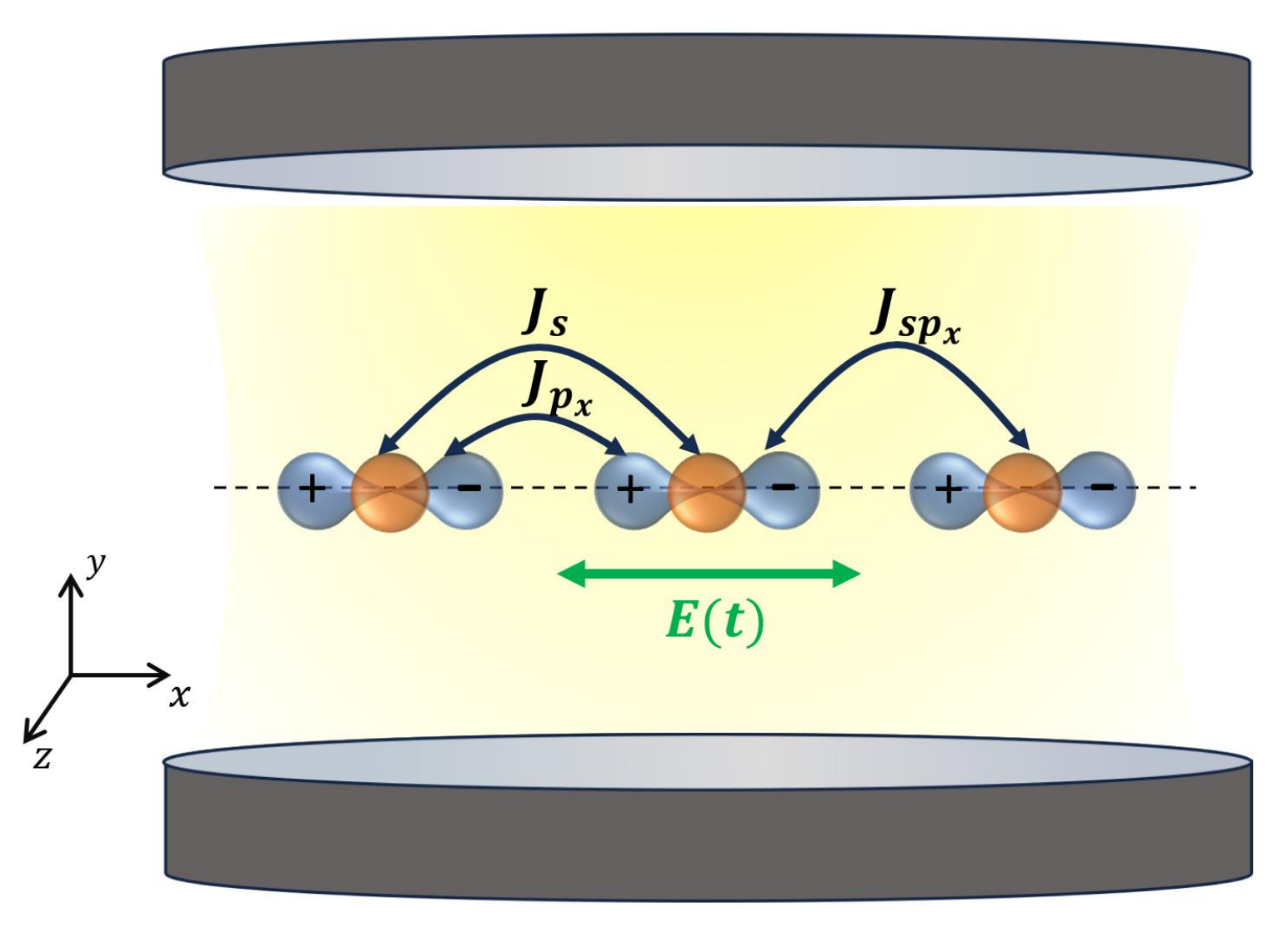}
		\caption{A one-dimensional electronic system consisting of two orbitals of opposite parities (\(s\) and \(p_x\)) at each site is centrally positioned in an optical cavity, which is shown by two large mirrors perpendicular to \(y\) axis. The intra-orbital \(J_\alpha\) (\(\alpha=s, p_x\)) and the nearest-neighbor inter-orbital  \(J_{sp_x}\) parameters describe the hoppings between the orbitals. The cavity's electromagnetic field is polarized along the lattice and propagates in the \(y\)-direction. }\label{fig:cavity-lattice}
	\end{figure}
	
	\section{Model and Method} \label{sec:model and method}
	We consider a one-dimensional lattice model hosting two orbitals with opposite parities at each lattice site: the \( s \) and \( p_x \) orbitals as shown in Fig. \ref{fig:cavity-lattice}. For simplicity, unless otherwise stated, we drop subindex \(x\) and the electron's spin is neglected. The Hamiltonian reads as
	\begin{equation}
		\hat{H}_{M} = \hat{H}_0 + \hat{H}_{\text{int}}\label{eq:1}
	\end{equation} 	
	
	The term \( \hat{H}_0 \), representing the kinetic energy, is 
	\begin{align}\nonumber
		\hat{H}_0 = &\sum_{i,\alpha} J_{\alpha} \hat{c}^\dagger_{i+1,\alpha} \hat{c}_{i,\alpha} + \sum_{i,\alpha} (D_{\alpha} - \mu) \hat{c}^\dagger_{i,\alpha} \hat{c}_{i,\alpha} \\
		&- J_{sp} \sum_i \left( \hat{c}^\dagger_{i+1,s} \hat{c}_{i,p} -\hat{c}^\dagger_{i-1,s} \hat{c}_{i,p}  \right)+\mathrm{h.c.}\label{eq:2}
	\end{align}
	\begin{figure}[t]
		\includegraphics[width=0.45\textwidth]{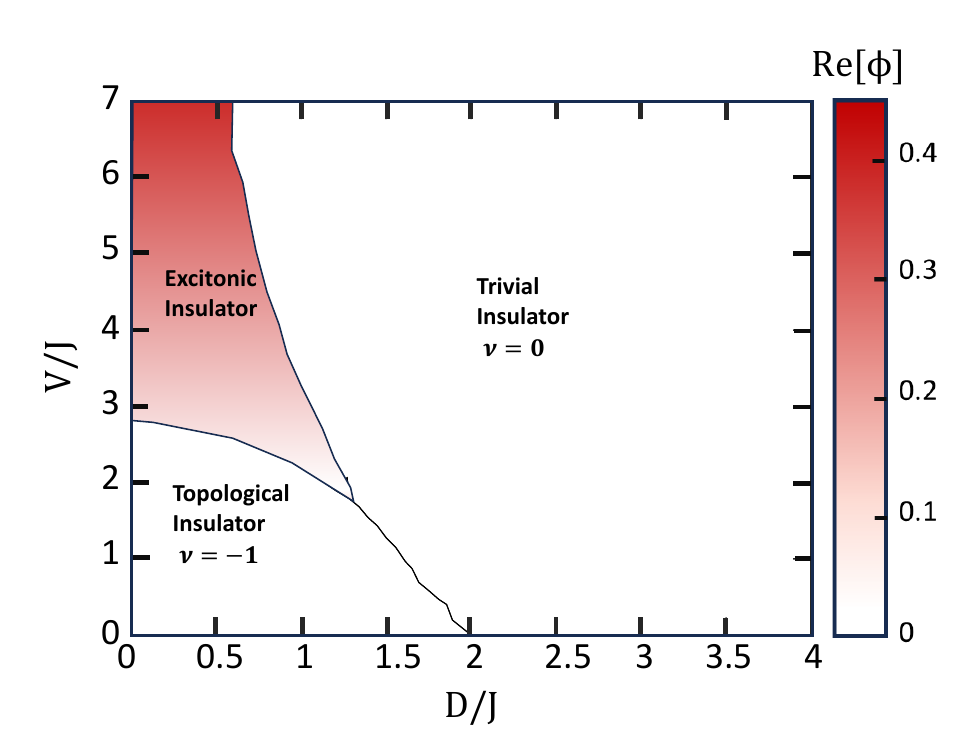}
		\caption{Phase diagram of the s–p chain model adapted (data is digitized) from \cite{khatibi2020excitonic} with permission. The real part of exciton order parameter \(\phi\) in the (D/J, V/J) plane and winding number $\nu$ are used to distinguish three phases. The colored area marks the excitonic insulator, while white regions denote topological (\(\nu=-1\)) and trivial (\(\nu=0\)) insulating phases, respectively.}\label{fig:phasediagram}
	\end{figure}
	
	In the above expression, \(\hat{c}^\dagger_{i,\alpha}\) (\(\hat{c}_{i,\alpha}\)) creates (annihilates) an electron at site \(i\) and in orbital \(\alpha \in \{s,p\} \). The parameter \( J_{\alpha} \) represents the intra-orbital hopping term, \( D_{\alpha} \) is the onsite orbital energy, \(\mu\) stands for the chemical potential, and \( J_{sp} \) indicates the inter-orbital hopping between adjacent orbitals of opposite parity within the lattice, i.e. \(J_{sp}(x)=-J_{sp}(-x)=-J_{sp}\). Fourier transformed to momentum space, we express \( \hat{H}_0 \) as:
	\begin{align}
		\hat{H}_0=\sum_{k,\alpha}\varepsilon_{k,\alpha}\hat{c}^\dagger_{k,\alpha}\hat{c}_{k,\alpha}+2i J_{sp}\sum_k \sin (ka) \hat{c}^\dagger_{k,s}\hat{c}_{k,p}+\mathrm{h.c.}\label{eq:3}
	\end{align}
	where \( \varepsilon_{k,\alpha} = 2J_\alpha \cos (ka) + D_\alpha - \mu \) with \(a\) as lattice constant. 
	
	The interaction term \( \hat{H}_{\text{int}} \) in (\ref{eq:1}) describes the local onsite interaction between electrons in the \(s\) and \(p\) orbitals:
	\begin{equation}
		\hat{H}_{\text{int}} = V \sum_i \hat{n}_{i,s} \hat{n}_{i,p}, \label{eq:4}
	\end{equation}
	where \( V \) indicates the strength of the Coulomb interaction, and \( \hat{n}_{i,\alpha} = \hat{c}^\dagger_{i,\alpha} \hat{c}_{i,\alpha} \) is the electron density operator. Employing a mean-field decomposition by introducing the exciton order parameter \( \phi = \langle \hat{c}^\dagger_{i,s} \hat{c}_{i,p} \rangle \) and the electron density \( n_{\alpha} = \langle \hat{c}^\dagger_{i,\alpha} \hat{c}_{i,\alpha} \rangle \), the interaction becomes
	\begin{align}
		\hat{H}_{\text{int}}^{\text{MF}} = V \sum_i \left( n_s \hat{n}_{i,p} + n_{p} \hat{n}_{i,s} - \phi^* \hat{c}_{i,s}^\dagger \hat{c}_{i,p} + \text{h.c.} \right).
		\label{eq:mean_field_interaction}
	\end{align}
	Using the Anderson pseudospin representation of orbitals \cite{anderson1958random}, the mean-field Hamiltonian reads as 
	\begin{align}
		H_M^{\text{MF}} = \sum_{k,\gamma} \hat{S}_k^\gamma B_k^\gamma,\label{eq:6}
	\end{align}
	where \( \hat{S}_k^\gamma = \frac{1}{2} \Psi_k^\dagger \sigma_\gamma \Psi_k \) represents a pseudospin operator with \( \Psi_k =( \hat{c}_{k,s}, \hat{c}_{k,p} )^{\text{T}} \) and \( \sigma_\gamma \) being the Pauli matrices for \( \gamma = 1, 2, 3 \) and the identity matrix for \( \gamma = 0 \). The components of the pseudomagnetic fields \( B_k^\gamma \) are computed as \cite{golevz2020nonlinear,murakami2017photoinduced,murakami2020collective,khatibi2020excitonic,kaneko2021bulk}:
	\begin{align}
		B^0_k&=V(n_s+n_{p})\\
		B^x_k&=-2V\mathrm{Re}[\phi]\\
		B^y_k&=-2V\mathrm{Im}[\phi]-4J_{sp}\sin (ka) \\
		B^z_k&=\varepsilon_{k,s}-\varepsilon_{k,{p}}+V(n_{p}-n_s).
	\end{align}
	
	Minimizing the free energy, the self-consistent equations are 
	\begin{align}
		&\phi=\frac{1}{N}\sum_k\frac{B_k^x+iB_k^y}{2B_k}[f(E_k^+,T)-f(E_k^-,T)], \label{eq:12-1}\\
		&n_s-n_{p}=\frac{1}{N}\sum_k\frac{B_k^z}{B_k}[f(E_k^+,T)-f(E_k^-,T)], \label{eq:12-2}\\
		&n_s+n_{p}=\frac{1}{N}\sum_k[f(E_k^+,T)+f(E_k^-,T)], \label{eq:12}
	\end{align}
	where $B_k=\sqrt{(B_k^x)^2+(B_k^y)^2+(B_k^z)^2}$, $E_k^{\pm}=(B_k^0\pm B_k)/2$ and $f(E^\pm_k,T)$ is the Fermi distribution function at temperature $T$.
	We set \( J_s = -J_{p} = -J \) and \( D_s = -D_{p} = D \) with \( J = 0.1 \mathrm{eV} \) as the unit of energy. The chemical potential \(\mu\) is chosen to ensure half-filling, \( n_s + n_{p} = 1 \), in (\ref{eq:12}). Fig. \ref{fig:phasediagram} reveals three distinct ground‐state phases—an excitonic insulator, a topological insulator, and a trivial band insulator—as functions of \(D/J\) and \(V/J\) \cite{khatibi2020excitonic}. Following up, we will extend this model to include the interaction with an optical cavity and explore the influence of these phases on the dispersion relations of the cavity modes.
	
	\section{Light - matter interaction in a  quantum cavity \label{sec:light-matter}}
	The model that we plan to study is shown schematically in Fig.~\ref{fig:cavity-lattice}, where a quantum s-p chain is placed in a cavity. The cavity consists of two parallel mirrors that reflect the electromagnetic field of incoming light, thereby defining the cavity modes. We confine our analysis to modes propagating in the \(y\) direction with polarization along the lattice. The cavity modes are described by 	\begin{align}
			\hat{H}_{pt} = \hbar \sum_q \omega(q) \hat{a}_q^\dagger \hat{a}_q \label{eq:14}
	\end{align}
	
	Here, \(\hat{a}_q^\dagger\)  (\(\hat{a}_q \)) creates (annihilates) a cavity photon mode carrying momentum \(q\), with energy dispersion \(\omega(q) = \sqrt{\omega_c^2+ c^2q^2 }\), in which \(\omega_c\) is the cavity’s fundamental frequency,  \(c\)  the speed of light, and \(\hbar\)  the reduced Planck constant. The cavity modes interact with the quantum system described by the following Hamiltonian:
	
	\begin{align}
		\hat{H}=\hat{H}_{MA}+\hat{H}_{EP}+\hat{H}_{pt},\label{eq:15}
	\end{align}
	where $\hat{H}_{MA}$ describes the mean-field Hamiltonian modified by a vector potential, $\hat{H}_{EP}$ describes the coupling between cavity modes and the electric dipoles. We briefly explain each term below. 
	
In the presence of an electromagnetic field, the cavity vector potential modifies the hopping amplitudes via the Peierls substitution \cite{li2020electromagnetic,sentef2020quantum,li2020manipulating,guerci2020superradiant,dmytruk2021gauge}, \(J_\alpha \rightarrow J_\alpha\exp\left[-\tfrac{ie}{\hbar}\int_{r_i}^{r_j}\vec{A}(r,t)\cdot dr\right]\) and \(J_{sp}\rightarrow J_{sp}\exp\left[-\tfrac{ie}{\hbar} \int_{r_i}^{r_j}\vec{A}(r,t)\cdot dr\right]\). For a nearly uniform \(\vec{A}(r,t)\) over lattice constant \(a\), the line integral reduces to \(\int_{r_i}^{r_j}\vec{A}(r,t)\cdot dr\simeq a\,\vec{A}_i(t)\), with \(\vec{A}_i(t)=\tfrac{A_0}{\sqrt{N}}(\hat a_i^\dagger+\hat a_i)\,\hat x\) denoting the cavity mode’s vector potential at site \(i\), \(N\) the number of unit cells and \(e\) the electron charge.  By defining a dimensionless parameter \(g\equiv eA_0a/\hbar\sqrt{N}\), $\hat{H}_{MA}$ is written as:
	\begin{align}\nonumber
		\hat{H}_{MA}&= \sum_{i,\alpha \in \{s,p\}} J_{\alpha} e^{-ig\left(\hat{a}_i^\dagger + \hat{a}_i\right)} \hat{c}^\dagger_{i+1,\alpha} \hat{c}_{i,\alpha}+ \sum_{i,\alpha \in \{s,p\}} (D_{\alpha}-\mu) \hat{c}^\dagger_{i,\alpha} \hat{c}_{i,\alpha}\\
		&- J_{sp}\sum_i\left( e^{-ig\left(\hat{a}_i^\dagger + \hat{a}_i\right)} \hat{c}^\dagger_{i+1,s} \hat{c}_{i,p} -e^{ig\left(\hat{a}_i^\dagger + \hat{a}_i\right)}\hat{c}^\dagger_{i-1,s} \hat{c}_{i,p}\right)\nonumber\\
		&+ V \sum_i \left( n_s \hat{n}_{i,p} + n_{p} \hat{n}_{i,s} - \phi^* \hat{c}_{i,s}^\dagger \hat{c}_{i,p}\right) + \text{h.c.}\label{eq:16}
	\end{align}
	In the thermodynamic limit (for large \(N\)), we use the Baker--Hausdorff formula
	\(\mathrm{exp}(\hat{X}+\hat{Y}) = \exp(\hat{X})\exp(\hat{Y})\exp (-[\hat{X},\hat{Y}]/2)\)
	to expand the above expression in terms of \(g\) up to linear term: 
	
	\begin{align}
		\hat{H}_{MA}\simeq\hat{H}_M^{MF}+\hat{H}_{LM}^{int},\label{eq:17}
	\end{align}
	where 
	\begin{align}
		\hat{H}_{LM}^{int}=\sum_{k,q}\sum_\nu\left(\hat{a}^\dagger_q+\hat{a}_{-q}\right)\mathcal{G}_\nu(k,q)\hat{\rho}_{k,\nu}(q)\label{eq:18}
	\end{align}
	with
	\(\hat{\rho}_{k,\nu}(q)=\Psi_k^\dagger \hat{\sigma}_\nu \Psi_{k+q}\) and  \(\check{\mathcal{G}}(k,q)\) is a diagonal matrix describing the electron-photon coupling strength, whose diagonal elements are 
	
	\begin{align}
		&\mathcal{G}_0(k,q)=-\frac{ig(J_s+J_{p})}{2}\left(e^{-ika}-e^{i(q+k)a}\right),\label{eq:19}\\
		&\mathcal{G}_1(k,q)=igJ_{sp}\left(\cos (ka) - \cos((k+q)a)\right),\label{eq:20}\\
		&\mathcal{G}_2(k,q)=-gJ_{sp}\left(\cos (ka) + \cos((k+q)a)\right),\label{eq:21}\\
		&\mathcal{G}_3(k,q)=-\frac{ig(J_s-J_{p})}{2}\left(e^{-ika}-e^{i(q+k)a}\right).\label{eq:22}
	\end{align}
	
	The hybridization $J_{sp}$ between $s$ and $p$ orbitals determines the coupling of photons with collective amplitude and phase modes via $\mathcal{G}_1(k,q)$ and $\mathcal{G}_2(k,q)$, respectively. Especially, at the limit of long-wave length $q\rightarrow0$, only the coupling to the phase mode survives. This observation is central to our discussions of the hybrid modes in Sec. \ref{sec:hybrid_modes}. 
	
	The second term in the Hamiltonian \eqref{eq:15} represents the interaction of the cavity electric field \(\hat{E} = -i (E_0/\sqrt{N}) \sum_i (\hat{a}_i^\dagger - \hat{a}_i)\) with the electric dipole at each site \(\hat{P} = ed_0 \sum_i (\hat{c}^\dagger_{i,s} \hat{c}_{i,p} + \text{h.c.})\) \cite{kaneko2021bulk,lenk2020collective}. Here, \(ed_0\) is the electric dipole amplitude between the \(s\) and \(p\) orbitals and \(E_0 = \omega_c/\hbar A_0\). By taking into account the interaction of the dipole and electric field in the form of \(\hat{E} \cdot \hat{P}\), \(\hat{H}_{EP}\) becomes:
	\begin{align}
		\hat{H}_{EP} = \frac{-ie E_0d_0}{\sqrt{N}} \sum_i  \left(\hat{a}_i^\dagger - \hat{a}_i\right) \left(\hat{c}^\dagger_{i,s} \hat{c}_{i,p} + \text{h.c.}\right).
	\end{align}
	
	Compared with the Hamiltonian \(\hat{H}_{MA}\), in our analysis we neglect \(\hat{H}_{EP}\) in \eqref{eq:15}. In fact, the coefficient of \(\hat{H}_{MA}\) with respect to \(\hat{H}_{EP}\) is proportional to \((e d_0 E_0 /\sqrt{N})/\tilde{J} g = \omega_c d_0/\tilde{J} a\), where \(\tilde{J} = J_\alpha, J_{sp}\). \(d_0\) is of the order of atomic size \(\sim 1 \text{\AA}\) and the lattice constant \(a\) is about \(\sim 4 \text{\AA}\) \cite{sunshine1985structure}. Additionally, we only consider \(\hbar\omega_c \ll \tilde{J} ~(\hbar\omega_c=0.1 J)\).
	
	Therefore, the effective light-matter interaction is described by \(\hat{H}=\hat{H}_M^{MF}+\hat{H}_{LM}^{int} + \hat{H}_{pt}\). We treat \(\hat{H}_{LM}^{int}\) perturbatively, and investigate the influence of material properties on photon propagation within the cavity. 
	
	\section{Spectroscopic Analysis of the Cavity Mode}\label{sec:Spectroscopic Analysis of the Cavity Mode}
	
	\subsection{Heterodyne Detection}
	The cavity modes can be studied using a heterodyne photodetector, an advanced optical instrument in quantum optics. The heterodyne photodetector employs two continuous light beams: one beam traverses the cavity, interacting with the cavity photon mode, while the other beam, serving as a local oscillator (LO), modulates the photons exiting the cavity. This modulation prepares the photons for precise measurement within the detector. Based on photoelectric \cite{vogel2006quantum} and input-output theory \cite{viviescas2003field}, the two-time correlation of photo-count in the detector can be related to correlations of the intra-cavity photon mode as: 
	
	\begin{align}
		&\frac{\overline{\hat{n}_q(t,\Delta t)\hat{n}_q(t^\prime,\Delta t)}-\overline{\hat{n}_q(t,\Delta t)}~ \overline{\hat{n}_q(t^\prime,\Delta t)}}{\overline{\hat{n}_q(t,\Delta t)}}\nonumber\\
		&\approx F \{e^{-i\omega_L t_{rel}}\langle\hat{a}_q^\dagger(t)\hat{a}_q(t^\prime)\rangle+\text{h.c.}\}, \label{eq:photodetector}
	\end{align}
	where \(\omega_L\) is the local oscillator frequency, \(F\) is the coefficient derived from the input-output and photoelectric calculations, and \(\hat{n}_q(t,\Delta t)\) represents the photon count operator at  momentum \(q\) in the time interval \(\left(t,t+\Delta t\right)\). \(\langle\hat{a}_q^\dagger(t)\hat{a}_q(t^\prime)\rangle\) is the intra-cavity photon correlation, which can be calculated from the photon Green's function. In the next subsection, we provide the expressions for intra-cavity photon dynamics.
	
	\subsection{Photon Green's function}\label{sec:Photonic spectroscopy within the cavity}	
	As discussed in the preceding section, the photon spectroscopy requires calculating the cavity photon Green's function given by [see appendix~ \eqref{app.photongf} for details]:
	
	\begin{align}
		\mathcal{D}(q,\omega)=\frac{\mathcal{D}_0(q,\omega)}{1-\mathcal{D}_0(q,\omega)\Pi(q,\omega)}.\label{eq:photonGF}
	\end{align}
	
	In this expression,  \(\mathcal{D}_0(q,\omega) = \frac{2\omega(q)}{(\omega + i0^+)^2 - \omega(q)^2}\) represents the bare photon Green's function. 
	\(\Pi(q,\omega)\)
	denotes the photon self-energy, which includes correlations from both screened electron-electron and bare electron-photon interactions. It consists of of two parts:
	\(\Pi(q,\omega) = \Pi^0(q,\omega) + \Pi^1(q,\omega)\), where
	
	\begin{widetext}
		\begin{align}
			&\Pi^0(q,\omega)=\frac{1}{\beta}\sum_{\mu,\nu}\sum_k\sum_{\omega^\prime}\mathcal{G}_\mu(k,q)\mathcal{G}_\nu(k+q,-q)\mathrm{Tr}\left[\check{G}^0(k,\omega^\prime)\sigma_\mu\check{G}^0(k+q,\omega^\prime+\omega)\sigma_\nu \right],\label{eq:25}\\
			&\Pi^1(q,\omega)=\frac{1}{\beta^2N}\sum_{\mu,\nu}\sum_{k,k^\prime}\sum_{\omega^\prime,\omega^{\prime\prime}}\sum_{\mu^\prime,\nu^\prime} \mathcal{G}_\nu(k+q,-q)\mathrm{Tr}\left[\check{G}^0(k^\prime,\omega^{\prime\prime})\sigma_{\nu^\prime} \check{G}^0(k^\prime+q,\omega^{\prime\prime}+\omega)\sigma_\nu\right]\nonumber\\	&~~~~~~~~~~~~~~~~~~~~~~~~~~~~~~~~~~~~~~~~~~~~~~~~~\times \check{V}_{\nu^\prime\mu^\prime}^{\text{eff}}(q,\omega)\mathrm{Tr}\left[\check{G}^0(k,\omega^\prime)\sigma_\mu\check{G}^0(k+q,\omega^\prime+\omega)\sigma_{\mu^\prime}\right]\mathcal{G}_\mu(k,q).\label{eq:26}
		\end{align} 
	\end{widetext} 
	In these equations, \(\beta = 1/T\) denotes the inverse temperature, \(\check{G}^0(k,\omega) = (\omega - \hat{H}^{MF}_M(k) + i0^+)^{-1}\) is the bare electron Green's function, and \(\mathrm{Tr}\) denotes the trace over the electronic states. The effective screened electron-electron interaction, \(\check{V}^{\text{eff}}(q,\omega) = (1 - \check{U}^0\chi^0(q,\omega))^{-1}\check{U}^0\), is computed in the RPA  whith \(\check{U}^0 = \frac{V}{2}\mathrm{diag}(1, -1, -1, -1)\) as the bare Coulomb potential and \(\chi^0(q,\omega)=\beta^{-1}\sum_k\sum_{\omega^\prime}\mathrm{Tr}[\sigma_\mu \check{G}^0(k+q,\omega^\prime+\omega)\sigma_\nu \check{G}^0(k,\omega^\prime)]\) is the bare polarization function of the electronic system. Now, using the \(\omega_L=0\) approximation \cite{jiang2008heterodyne} in Eq. \eqref{eq:photodetector}, the poles of  \(\mathcal{D}(q,\omega)\), representing the photon energy dispersion within an optical cavity, constitute the heterodyne photodetector response. Next, we employ the above expressions to explore how the cavity photon's energy is influenced when coupled to the electronic degrees of freedom of the one-dimensional s-p chain.
	
	\begin{figure*}[t]
		\includegraphics[width=0.9\textwidth]{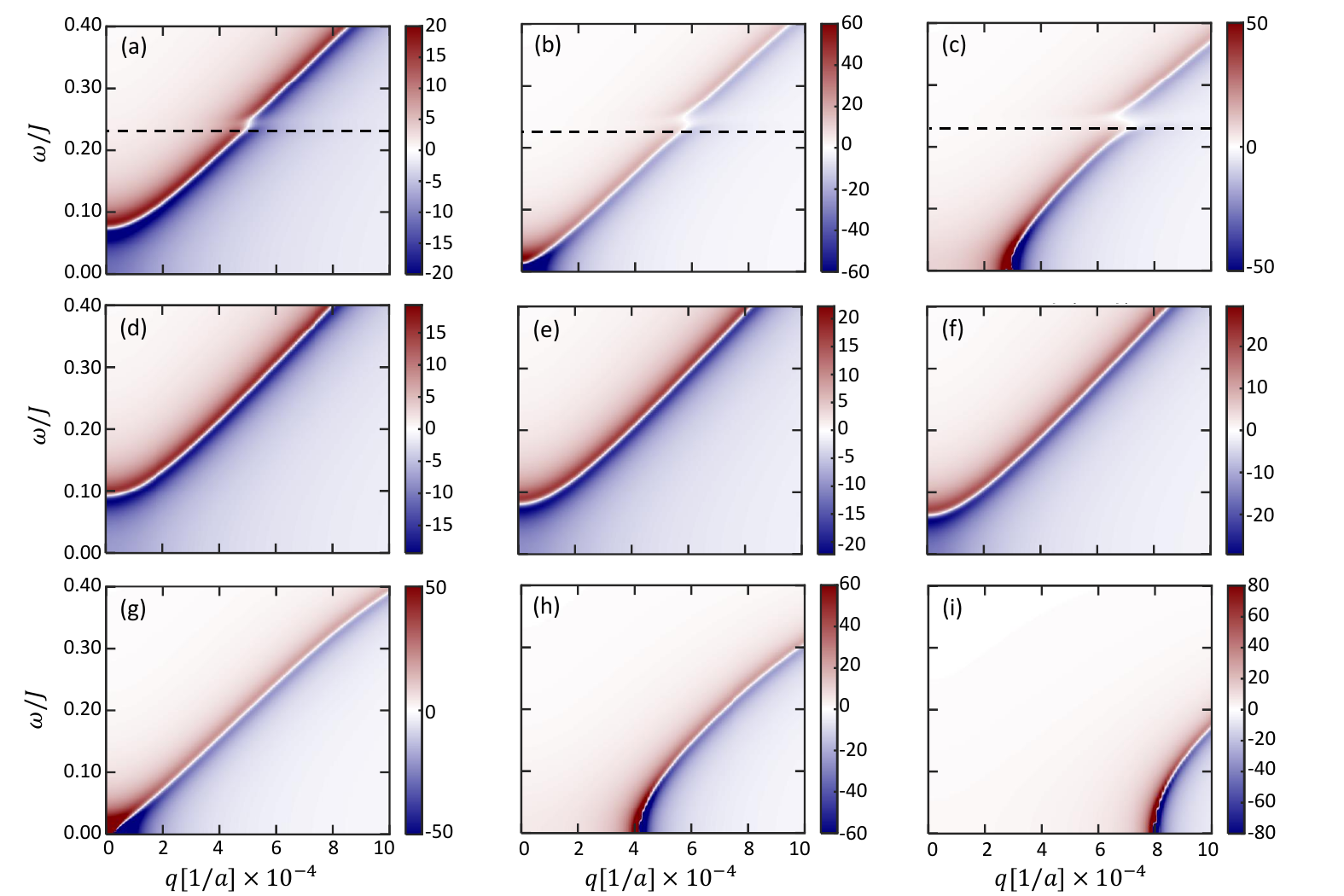}
		\caption{The heat map of \(\Re[\mathcal{D}(q,\omega)]\) is shown. We adopt natural units such that \(\hbar=6.6\times10^{-16} \mathrm{eV.s}\) and \(c=3\times10^8 \mathrm{ m/s}\) and set the cavity photon fundamental frequency to \(\hbar\omega_c/J=0.1\). The other parameters are fixed as \(J=0.1\mathrm{eV}\) and lattice constant \(a=4 \text{\AA}\). Panels (a,b,c) correspond to the excitonic insulator phase (parameters: \(V/J=2.5, D/J=1.1, J_{sp}/J=0.5, \phi=0.116\)), (d,e,f) to the trivial insulator phase (parameters: \(V/J=2.5, D/J=2, J_{sp}/J=0.5, \phi=0\)), and (g,h,i) to the topological insulator phase (parameters: \(V/J=1, D/J=1.1, J_{sp}/J=0.5, \phi=0\)). In each row, from left to right, panels correspond to the strength of the light-matter interaction \(g=0.2\), \(g=0.3\) and \(g=0.4\). In panels (a–c), the dashed black line marks the excitonic phase mode energy, approximately \(\omega_{\mathrm{phase}}/J=0.23\).} 
		\label{fig:Dpt}
	\end{figure*}

	\section{collective hybrid modes \label{sec:hybrid_modes}}
	
	In Fig.~\ref{fig:Dpt} we show \(\Re[\mathcal{D}(q,\omega)]\) to study the spectrum of the cavity photon mode when coupled to the ground state of the s-p chain in different phases (for completeness  -\(\Im[\mathcal{D}(q,\omega)]\) is shown in appendix~\ref{app.imagGF}). 
	
	{\it Excitonic insulator}-- 
	Figs. \ref{fig:Dpt}(a,b,c) correspond to the excitonic insulator phase; the  parameters are \(V/J=2.5\), \( D/J=1.1\) and \(J_{sp}/J=0.5\) yielding \(\phi=0.116\). We examine the photon spectrum by varying electron-photon interaction strengths \(g\). In the absence of light-matter interaction, the bare photon branch follows \(\omega(q)=\sqrt{\omega_c^2+c^2q^2}\),  with \(\hbar\omega_c/J=0.1\), while the excitonic phase mode exhibits a gap at \(q=0\) of approximately \(\omega_{\mathrm{phase}}/J \approx 0.23\), due to \(U(1)\) symmetry breaking induced by the finite \(J_{sp}\) (see appendix~ \ref{app.collective mode} for details). The photon branch is much more dispersive than the phase mode. Hence, we expect hybridization at small momenta, where the phase mode is almost dispersionless shown by a dashed line.  The finite interaction strength \(g\) in \eqref{eq:18} hybridizes the modes, resulting in an avoiding band crossing. As the light-matter interaction strength increases, both branches move toward lower energies. At a critical coupling \(g_c\approx0.3\), the lower branch softens completely to zero energy at \(q=0\). For \(g>g_c\), the softening instead occurs at a finite critical momentum \(q_c\), beyond which the mode recovers a nonzero dispersion. Physically photon softening signals the onset of photon condensation; to make this explicit, within Landau–Ginzburg theory the static free‐energy for the photon quadrature \(X_q\equiv\langle \hat{a}_q +\hat{a}^\dagger_{-q}\rangle \)  takes the form \(F(X_q)=\frac{1}{2}\alpha X_q^2+\frac{1}{4}\beta X_q^4\), with \(\beta>0\) and \(\alpha\propto 1+2\Pi(q,0)/[\hbar\omega(q)]\). Above threshold (\(\alpha>0\)) the minimum lies at \(X_q=0\), but once \(\alpha\) vanishes at the critical wavevector \(q_c\) the quartic term stabilizes a finite solution \(X_q=-2\alpha/\beta>0\). At $q=0$, we found $\alpha<0$ for $g>g_c\approx 0.3$. Although we established a condensed phase, its physical implication remains unclear. A tentative hypothesis is that photons may be squeezed into coherent states. However, the true nature of the condensed phase \cite{mazza2019superradiant, andolina2019cavity, Mazza2023} and its physical significance shall be explored in future studies.

{\it Trivial insulator}-- Next, we consider a regime of parameters where the electronic system is a trivial insulator. For that, we use \(V/J=2.5\),  \(D/J=2\) and \(J_{sp}/J=0.5\), where \(\phi=0\), and hence no exciton condensation. The results of coupling to cavity photon modes are shown in Fig.~\ref{fig:Dpt}(d,e,f). As seen here, the cavity mode remains intact and there in no hybridization due to the absence of electronic excitations within the insulator gap. Even by increasing the coupling, the photon mode remains unchanged as if there is no insulator medium around. The spectrum is clearly distinct from the excitonic insulator presented in Figs. \ref{fig:Dpt}(a,b,c). 
	
	{\it Topological insulator}-- As shown in Ref.\cite{khatibi2020excitonic} and schematically in Fig. \ref{fig:phasediagram}, the one-dimensional s-p chain allows for a topological insulator phase in a wide range of parameters. For our purposes we set the parameters as \(V/J=1\), \(D/J=1.1\) and \(J_{sp}/J=0.5\), for which the excitonic order parameter \(\phi=0\). Nevertheless, the winding number is $\nu=1$ and hence the model in topologically nontrivial. Upon coupling to the cavity modes, the dispersion of photon mode is shown in Fig.~\ref{fig:Dpt}(g,h,i). At small \(g<0.1\), the photon mode remains unchanged. As \(g\) increases, the entire dispersion shifts to lower energy, and at a critical coupling \(g_c\approx0.1\), the photon energy at \(q=0\) softens to zero. Beyond \(g_c\), the photon mode softens to zero energy at a finite momentum \(q_c\). Again, since the ground state is free of any excitonic condensation, no hybridization is observed.        
	
	From the above observations, we conclude that the presence of an excitonic insulator with excitonic condensate -- as opposed to trivial and topological insulators -- has significant effects on the cavity photon mode, which is inferred from the photon self-energy \(\Pi(q,\omega)\).

	From the electron-photon interaction Hamiltonian, in our system with \(J_s = -J_p\) we have \(\mathcal{G}_0(q,\omega) = 0\) according to \eqref{eq:19}. Additionally \(\mathcal{G}_1(q,\omega)\) and \(\mathcal{G}_2(q,\omega)\) describe couplings to the oscillations along the Higgs and Goldstone modes of the excitonic insulator, respectively, and \(\mathcal{G}_3(q,\omega)\) couples to the fluctuations of the charge density. In an excitonic insulator, fluctuations of the collective modes induce light-matter coupling and affect the cavity photon energy. This facilitates energy transfer between matter and light, resulting in Rabi oscillations \cite{scully1997quantum,haroche1989cavity} between phase and photon modes and hence an avoiding band crossing. For a topological insulator when perturbed by light, the single particle excitations across the gap produce charge fluctuations, which modify the cavity photon energy. For trivial insulator, due to the large band gap compared to the topological insulator charge fluctuations are very weak and thus the photon mode is less affected by the material.

	\begin{figure*}[t!]
		\includegraphics[width=1\textwidth]{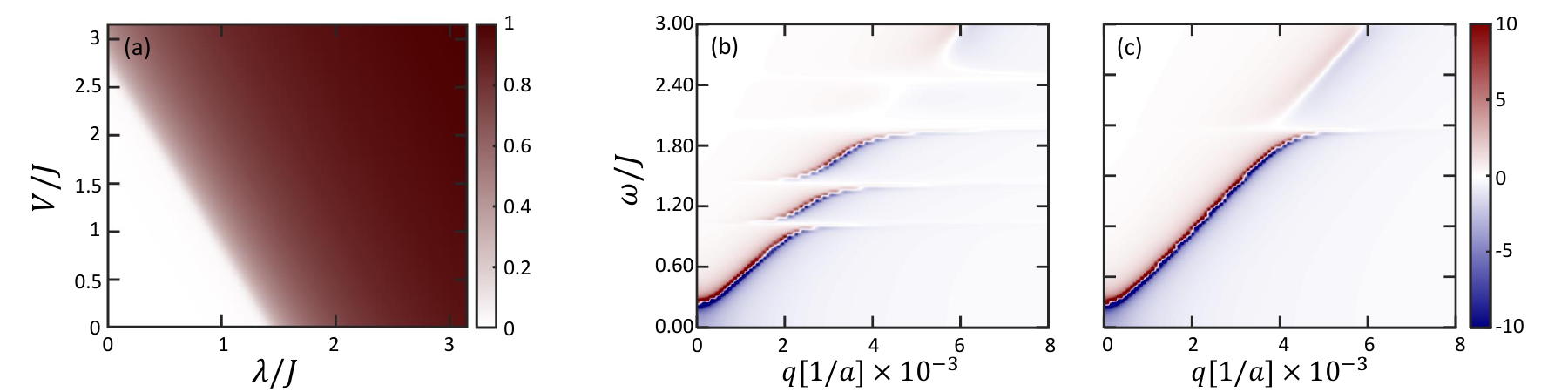}
		\caption{(a) Density plot of the excitonic order parameter \(\phi\) in the \(\lambda/J-V/J\) plan. Heat maps of \(\Re[\mathcal{D}(q,\omega)]\) in (b) Coulomb-driven (parameters: \( V/J=2.8\), \(D/J=0.5\), \(J_{sp}/J=0.5\), \(\phi=0.1755\), \(\lambda/J=0.1\) ) and (c) phonon-driven (parameters: \( V/J=0.1\), \(D/J=0.5\), \(J_{sp}/J=0.5\), \(\lambda/J=1.4\), \(\phi=0.075\)) excitonic-insulator phases. In both regimes the single-particle gap is set to \(E_g/J\approx 2\), the light-matter interaction to \(g=0.2\), phonon frequency to \(\omega_{\mathrm{pn}}/J=0.1\) and the cavity fundamental frequency to \(\hbar\omega_c/J=0.3\) (\(J=0.1\) and \(a=4 \text{\AA}\)).} \label{fig:photonspectrumandphasespace-eph}
	\end{figure*}
	
\section{Coulomb- and Phonon-Mediated Pairing in the Excitonic Insulator} \label{sec:phonon} 
Excitonic pairing arises both from direct electron–electron Coulomb interactions and from coupling to lattice vibrations. To capture both mechanisms, we extend Hamiltonian  \eqref{eq:1} by adding a Holstein-type term that couples the electronic density to an optical phonon mode. This isolates the Coulomb-dominated and phonon-dominated excitonic-insulator phases and reveals how each regime imprints on the cavity-photon spectrum. We augment \(\hat{H}_{\mathrm{int}}\) by adding the electron–phonon coupling term \(\hat{H}_{\mathrm{e-pn}}\) \cite{watanabe2015charge,Zenker2013,phan2014linear,zenker2014fate,murakami2020collective,golevz2020nonlinear,kim2020phononic}:

	\begin{align}
		\hat{H}_{\mathrm{e-pn}} = \zeta \sum_i \left(\hat{b}_i^\dagger + \hat{b}_i\right) \left(\hat{c}^\dagger_{is} \hat{c}_{i,p} + \mathrm{h.c.}\right),
	\end{align}
	and the phonon Hamiltonian is given by:
	
	\begin{align}
		\hat{H}_{\text{pn}}=\omega_{\text{pn}}\sum_i\hat{b}_i^\dagger\hat{b}_i.
	\end{align} 
	
	Here, \(\hat{b}_i\) (\(\hat{b}_i^\dagger\)) denotes the phonon annihilation (creation) operator, \(\omega_{\text{pn}}\) represents the phonon energy, \(\zeta\) is the electron-phonon coupling constant, and the effective electron-phonon coupling is defined as \(\lambda \equiv 2\zeta^2/\omega_{\text{pn}}\). Treating the interaction term in the mean-field approximation, we define \(X = \langle \hat{b}_i^\dagger + \hat{b}_i \rangle\), which relates to the exciton order parameter by \(X = -4\zeta \mathrm{Re}\left[\phi\right]/\omega_{\text{pn}}\). Consequently, in the pseudospin representation of the total mean-field Hamiltonian \eqref{eq:6}, the electron–phonon interaction modifies only the x-component of the pseudomagnetic field such that
		\(B^x_k=-2\mathrm{Re}[\phi]\left(V+2\lambda\right)\), while all other components remain unchanged.
	Solving the mean-field equations (\ref{eq:12-1}-\ref{eq:12}) allows us to map out the ground state phase diagram as shown in Fig~\ref{fig:photonspectrumandphasespace-eph}~(a). 

	Upon coupling the s-p chain to the optical modes of cavity and using the same procedures outlined in Sec. \eqref{sec:Photonic spectroscopy within the cavity}, we compute the Green function of photons. The effective electron–electron interaction now includes both the intrinsic Coulomb repulsion and a phonon-mediated term arising from electron–phonon coupling, resulting in the matrix \(\check{U}^0 = \mathrm{diag}(\frac{V}{2}, -\frac{V}{2}+\lambda^2 \mathcal{D}^0_{\text{pn}}(q,\omega), -\frac{V}{2}, -\frac{V}{2})\), where the bare phonon propagator is given by \(\mathcal{D}^0_{\text{pn}}(q,\omega) = 2\omega_{\text{pn}}/[(\omega + i0^+)^2 - \omega_{\text{pn}}^2\)]. Figures \ref{fig:photonspectrumandphasespace-eph} (b) and (c) display the cavity-photon energy spectra for the Coulomb-dominated and phonon-dominated excitonic-insulator phases, respectively. In both cases, we fix the single-particle gap to \(E_g/J \approx 2\), the phonon frequency to \(\omega_{\mathrm{pn}}/J=0.1\), and the light–matter coupling to \(g=0.2\). In the Coulomb-dominated regime, the collective phase mode sits just below the single-particle gap (see Fig.~\ref{fig:collective mode} (e) in Appendix~\ref{app.collective mode}); tuning the photon fundamental frequency \(\omega_c/J=0.3\) beneath this mode produces strong hybridization and an avoided crossing, splitting the photonic branch at resonance. By contrast, when electron–phonon coupling prevails the phase mode is pushed into the electron–hole continuum \cite{murakami2020collective,golevz2020nonlinear}, so the photon instead hybridizes with a broad continuum of excitations. This washes out the sharp splitting and restores a featureless, normal-insulator–like response. Consequently, measuring the cavity-photon spectrum provides a clear diagnostic: the observation of a split photonic branch below the single-particle gap unambiguously signals the Coulomb-dominated excitonic-insulator phase, whereas its absence indicates the phonon-dominated regime.

	\section{Conclusions}\label{sec:conclusion}
	Recent optical and transport measurements present controversial conclusions on the nature of the insulating ground states of some of transition metal chalcogenides, the famous one is \(\mathrm{Ta_2NiSe_5}\). While some of measurements are inclined to claim that the gap opening is caused by the condensation of excitons \cite{golevz2022unveiling,bretscher2021imaging,kim2021direct,mazza2020nature}, there are also evidences that the ground state might be a trivial insulator due to structural phase transition \cite{baldini2023spontaneous,windgatter2021common,kim2020phononic}. A large body of works used optical pulses in the classical regime to examine the nature of the insulator phase \cite{golevz2022unveiling,bretscher2021imaging,baldini2023spontaneous,golevz2020nonlinear,murakami2020collective,murakami2017photoinduced}. 
	
	Motivated by these observations, in this paper, we explore the response of an excitonic insulator using quantum nature of light in a quantum cavity. We posed the following question: given a model with both excitonic (driven by electronic Coulomb interaction) and trivial ground states, how is the dispersion of photon mode in the cavity modified? The chief goal has been to answer this question. We used a one-dimensional s-p chain lattice model, whose phase diagram has three distinct phases: excitonic insulator, topological insulator, and trivial band insulator. The ground state may also include phonon-mediated excitons. In an optical cavity we investigated the interplay between light and matter in different phases, focusing on the impact of excitonic condensation on cavity photon modes. Our results show that the response of an excitonic ground state  formed by a coherent condensation of excitons significantly differs from the insulator phases with no condensation or even with excitons created by phonons. For the coherent exciton phase,  the light-matter coupling leads to entangled electron - photon states and an avoiding band crossing is observed in the collective excitations. This singles out the excitonic insulators from trivial and topological insulators, where the photon dispersion essentially remains unchanged and the charge fluctuations between two bands result only in changing the photon's intensity and lowering its energy. These changes are more pronounced in topological insulators than in trivial insulators due to the large band gap in the latter. 
	
Building on our theoretical predictions, we suggest heterodyne photodetection \cite{vogel2006quantum,loudon2000quantum,lysne2023quantum,grynberg2010introduction} as a viable method to experimentally probe the interplay between light and matter within the cavity. By detecting the interference between a stable reference light and photons modified by the different phases of the system in the cavity, this technique can map the spectral signatures unique to excitonic condensation. Furthermore, it can discern whether the exciton condensation within the system is driven by electronic interactions or mediated by phonons. These observations can provide experimental insights into the distinct collective excitations predicted in our work.

	\section{Acknowledgement}
	The authors would like to thank Sharif University of Technology for supports.  We also thank Microsoft's Copilot for revising several sentences to enhance clarity.


\begin{thebibliography}{65}%
\makeatletter
\providecommand \@ifxundefined [1]{%
 \@ifx{#1\undefined}
}%
\providecommand \@ifnum [1]{%
 \ifnum #1\expandafter \@firstoftwo
 \else \expandafter \@secondoftwo
 \fi
}%
\providecommand \@ifx [1]{%
 \ifx #1\expandafter \@firstoftwo
 \else \expandafter \@secondoftwo
 \fi
}%
\providecommand \natexlab [1]{#1}%
\providecommand \enquote  [1]{``#1''}%
\providecommand \bibnamefont  [1]{#1}%
\providecommand \bibfnamefont [1]{#1}%
\providecommand \citenamefont [1]{#1}%
\providecommand \href@noop [0]{\@secondoftwo}%
\providecommand \href [0]{\begingroup \@sanitize@url \@href}%
\providecommand \@href[1]{\@@startlink{#1}\@@href}%
\providecommand \@@href[1]{\endgroup#1\@@endlink}%
\providecommand \@sanitize@url [0]{\catcode `\\12\catcode `\$12\catcode
  `\&12\catcode `\#12\catcode `\^12\catcode `\_12\catcode `\%12\relax}%
\providecommand \@@startlink[1]{}%
\providecommand \@@endlink[0]{}%
\providecommand \url  [0]{\begingroup\@sanitize@url \@url }%
\providecommand \@url [1]{\endgroup\@href {#1}{\urlprefix }}%
\providecommand \urlprefix  [0]{URL }%
\providecommand \Eprint [0]{\href }%
\providecommand \doibase [0]{https://doi.org/}%
\providecommand \selectlanguage [0]{\@gobble}%
\providecommand \bibinfo  [0]{\@secondoftwo}%
\providecommand \bibfield  [0]{\@secondoftwo}%
\providecommand \translation [1]{[#1]}%
\providecommand \BibitemOpen [0]{}%
\providecommand \bibitemStop [0]{}%
\providecommand \bibitemNoStop [0]{.\EOS\space}%
\providecommand \EOS [0]{\spacefactor3000\relax}%
\providecommand \BibitemShut  [1]{\csname bibitem#1\endcsname}%
\let\auto@bib@innerbib\@empty
\bibitem [{\citenamefont {J{\'e}rome}\ \emph {et~al.}(1967)\citenamefont
  {J{\'e}rome}, \citenamefont {Rice},\ and\ \citenamefont
  {Kohn}}]{jerome1967excitonic}%
  \BibitemOpen
  \bibfield  {author} {\bibinfo {author} {\bibfnamefont {D.}~\bibnamefont
  {J{\'e}rome}}, \bibinfo {author} {\bibfnamefont {T.}~\bibnamefont {Rice}},\
  and\ \bibinfo {author} {\bibfnamefont {W.}~\bibnamefont {Kohn}},\ }\bibfield
  {title} {\bibinfo {title} {Excitonic insulator},\ }\href@noop {} {\bibfield
  {journal} {\bibinfo  {journal} {Physical Review}\ }\textbf {\bibinfo {volume}
  {158}},\ \bibinfo {pages} {462} (\bibinfo {year} {1967})}\BibitemShut
  {NoStop}%
\bibitem [{\citenamefont {Kohn}(1967)}]{kohn1967excitonic}%
  \BibitemOpen
  \bibfield  {author} {\bibinfo {author} {\bibfnamefont {W.}~\bibnamefont
  {Kohn}},\ }\bibfield  {title} {\bibinfo {title} {Excitonic phases},\
  }\href@noop {} {\bibfield  {journal} {\bibinfo  {journal} {Physical Review
  Letters}\ }\textbf {\bibinfo {volume} {19}},\ \bibinfo {pages} {439}
  (\bibinfo {year} {1967})}\BibitemShut {NoStop}%
\bibitem [{\citenamefont {Keldysh}\ and\ \citenamefont
  {Kozlov}(2024)}]{keldysh2024collective}%
  \BibitemOpen
  \bibfield  {author} {\bibinfo {author} {\bibfnamefont {L.}~\bibnamefont
  {Keldysh}}\ and\ \bibinfo {author} {\bibfnamefont {A.}~\bibnamefont
  {Kozlov}},\ }\bibfield  {title} {\bibinfo {title} {Collective properties of
  excitons in semiconductors},\ }in\ \href@noop {} {\emph {\bibinfo {booktitle}
  {SELECTED PAPERS OF LEONID V KELDYSH}}}\ (\bibinfo  {publisher} {World
  Scientific},\ \bibinfo {year} {2024})\ pp.\ \bibinfo {pages}
  {79--86}\BibitemShut {NoStop}%
\bibitem [{\citenamefont {Halperin}\ and\ \citenamefont
  {Rice}(1968{\natexlab{a}})}]{halperin1968excitonic}%
  \BibitemOpen
  \bibfield  {author} {\bibinfo {author} {\bibfnamefont {B.}~\bibnamefont
  {Halperin}}\ and\ \bibinfo {author} {\bibfnamefont {T.}~\bibnamefont
  {Rice}},\ }\bibfield  {title} {\bibinfo {title} {The excitonic state at the
  semiconductor-semimetal transition},\ }in\ \href@noop {} {\emph {\bibinfo
  {booktitle} {Solid State Physics}}},\ Vol.~\bibinfo {volume} {21}\ (\bibinfo
  {publisher} {Elsevier},\ \bibinfo {year} {1968})\ pp.\ \bibinfo {pages}
  {115--192}\BibitemShut {NoStop}%
\bibitem [{\citenamefont {Halperin}\ and\ \citenamefont
  {Rice}(1968{\natexlab{b}})}]{halperin1968possible}%
  \BibitemOpen
  \bibfield  {author} {\bibinfo {author} {\bibfnamefont {B.}~\bibnamefont
  {Halperin}}\ and\ \bibinfo {author} {\bibfnamefont {T.}~\bibnamefont
  {Rice}},\ }\bibfield  {title} {\bibinfo {title} {Possible anomalies at a
  semimetal-semiconductor transistion},\ }\href@noop {} {\bibfield  {journal}
  {\bibinfo  {journal} {Reviews of Modern Physics}\ }\textbf {\bibinfo {volume}
  {40}},\ \bibinfo {pages} {755} (\bibinfo {year}
  {1968}{\natexlab{b}})}\BibitemShut {NoStop}%
\bibitem [{\citenamefont {Butov}\ \emph {et~al.}(2002)\citenamefont {Butov},
  \citenamefont {Gossard},\ and\ \citenamefont
  {Chemla}}]{butov2002macroscopically}%
  \BibitemOpen
  \bibfield  {author} {\bibinfo {author} {\bibfnamefont {L.}~\bibnamefont
  {Butov}}, \bibinfo {author} {\bibfnamefont {A.}~\bibnamefont {Gossard}},\
  and\ \bibinfo {author} {\bibfnamefont {D.}~\bibnamefont {Chemla}},\
  }\bibfield  {title} {\bibinfo {title} {Macroscopically ordered state in an
  exciton system},\ }\href@noop {} {\bibfield  {journal} {\bibinfo  {journal}
  {Nature}\ }\textbf {\bibinfo {volume} {418}},\ \bibinfo {pages} {751}
  (\bibinfo {year} {2002})}\BibitemShut {NoStop}%
\bibitem [{\citenamefont {Eisenstein}\ and\ \citenamefont
  {MacDonald}(2004)}]{eisenstein2004bose}%
  \BibitemOpen
  \bibfield  {author} {\bibinfo {author} {\bibfnamefont {J.}~\bibnamefont
  {Eisenstein}}\ and\ \bibinfo {author} {\bibfnamefont {A.~H.}\ \bibnamefont
  {MacDonald}},\ }\bibfield  {title} {\bibinfo {title} {Bose--einstein
  condensation of excitons in bilayer electron systems},\ }\href@noop {}
  {\bibfield  {journal} {\bibinfo  {journal} {Nature}\ }\textbf {\bibinfo
  {volume} {432}},\ \bibinfo {pages} {691} (\bibinfo {year}
  {2004})}\BibitemShut {NoStop}%
\bibitem [{\citenamefont {Wang}\ \emph {et~al.}(2019)\citenamefont {Wang},
  \citenamefont {Rhodes}, \citenamefont {Watanabe}, \citenamefont {Taniguchi},
  \citenamefont {Hone}, \citenamefont {Shan},\ and\ \citenamefont
  {Mak}}]{wang2019evidence}%
  \BibitemOpen
  \bibfield  {author} {\bibinfo {author} {\bibfnamefont {Z.}~\bibnamefont
  {Wang}}, \bibinfo {author} {\bibfnamefont {D.~A.}\ \bibnamefont {Rhodes}},
  \bibinfo {author} {\bibfnamefont {K.}~\bibnamefont {Watanabe}}, \bibinfo
  {author} {\bibfnamefont {T.}~\bibnamefont {Taniguchi}}, \bibinfo {author}
  {\bibfnamefont {J.~C.}\ \bibnamefont {Hone}}, \bibinfo {author}
  {\bibfnamefont {J.}~\bibnamefont {Shan}},\ and\ \bibinfo {author}
  {\bibfnamefont {K.~F.}\ \bibnamefont {Mak}},\ }\bibfield  {title} {\bibinfo
  {title} {Evidence of high-temperature exciton condensation in two-dimensional
  atomic double layers},\ }\href@noop {} {\bibfield  {journal} {\bibinfo
  {journal} {Nature}\ }\textbf {\bibinfo {volume} {574}},\ \bibinfo {pages}
  {76} (\bibinfo {year} {2019})}\BibitemShut {NoStop}%
\bibitem [{\citenamefont {Nandi}\ \emph {et~al.}(2012)\citenamefont {Nandi},
  \citenamefont {Finck}, \citenamefont {Eisenstein}, \citenamefont {Pfeiffer},\
  and\ \citenamefont {West}}]{nandi2012exciton}%
  \BibitemOpen
  \bibfield  {author} {\bibinfo {author} {\bibfnamefont {D.}~\bibnamefont
  {Nandi}}, \bibinfo {author} {\bibfnamefont {A.}~\bibnamefont {Finck}},
  \bibinfo {author} {\bibfnamefont {J.}~\bibnamefont {Eisenstein}}, \bibinfo
  {author} {\bibfnamefont {L.}~\bibnamefont {Pfeiffer}},\ and\ \bibinfo
  {author} {\bibfnamefont {K.}~\bibnamefont {West}},\ }\bibfield  {title}
  {\bibinfo {title} {Exciton condensation and perfect coulomb drag},\
  }\href@noop {} {\bibfield  {journal} {\bibinfo  {journal} {Nature}\ }\textbf
  {\bibinfo {volume} {488}},\ \bibinfo {pages} {481} (\bibinfo {year}
  {2012})}\BibitemShut {NoStop}%
\bibitem [{\citenamefont {Sunshine}\ and\ \citenamefont
  {Ibers}(1985)}]{sunshine1985structure}%
  \BibitemOpen
  \bibfield  {author} {\bibinfo {author} {\bibfnamefont {S.~A.}\ \bibnamefont
  {Sunshine}}\ and\ \bibinfo {author} {\bibfnamefont {J.~A.}\ \bibnamefont
  {Ibers}},\ }\bibfield  {title} {\bibinfo {title} {Structure and physical
  properties of the new layered ternary chalcogenides tantalum nickel sulfide
  (ta2nis5) and tantalum nickel selenide (ta2nise5)},\ }\href@noop {}
  {\bibfield  {journal} {\bibinfo  {journal} {Inorganic Chemistry}\ }\textbf
  {\bibinfo {volume} {24}},\ \bibinfo {pages} {3611} (\bibinfo {year}
  {1985})}\BibitemShut {NoStop}%
\bibitem [{\citenamefont {Wakisaka}\ \emph {et~al.}(2009)\citenamefont
  {Wakisaka}, \citenamefont {Sudayama}, \citenamefont {Takubo}, \citenamefont
  {Mizokawa}, \citenamefont {Arita}, \citenamefont {Namatame}, \citenamefont
  {Taniguchi}, \citenamefont {Katayama}, \citenamefont {Nohara},\ and\
  \citenamefont {Takagi}}]{wakisaka2009excitonic}%
  \BibitemOpen
  \bibfield  {author} {\bibinfo {author} {\bibfnamefont {Y.}~\bibnamefont
  {Wakisaka}}, \bibinfo {author} {\bibfnamefont {T.}~\bibnamefont {Sudayama}},
  \bibinfo {author} {\bibfnamefont {K.}~\bibnamefont {Takubo}}, \bibinfo
  {author} {\bibfnamefont {T.}~\bibnamefont {Mizokawa}}, \bibinfo {author}
  {\bibfnamefont {M.}~\bibnamefont {Arita}}, \bibinfo {author} {\bibfnamefont
  {H.}~\bibnamefont {Namatame}}, \bibinfo {author} {\bibfnamefont
  {M.}~\bibnamefont {Taniguchi}}, \bibinfo {author} {\bibfnamefont
  {N.}~\bibnamefont {Katayama}}, \bibinfo {author} {\bibfnamefont
  {M.}~\bibnamefont {Nohara}},\ and\ \bibinfo {author} {\bibfnamefont
  {H.}~\bibnamefont {Takagi}},\ }\bibfield  {title} {\bibinfo {title}
  {Excitonic insulator state in ta 2 nise 5 probed by photoemission
  spectroscopy},\ }\href@noop {} {\bibfield  {journal} {\bibinfo  {journal}
  {Physical review letters}\ }\textbf {\bibinfo {volume} {103}},\ \bibinfo
  {pages} {026402} (\bibinfo {year} {2009})}\BibitemShut {NoStop}%
\bibitem [{\citenamefont {Kaneko}\ \emph {et~al.}(2012)\citenamefont {Kaneko},
  \citenamefont {Seki},\ and\ \citenamefont {Ohta}}]{kaneko2012excitonic}%
  \BibitemOpen
  \bibfield  {author} {\bibinfo {author} {\bibfnamefont {T.}~\bibnamefont
  {Kaneko}}, \bibinfo {author} {\bibfnamefont {K.}~\bibnamefont {Seki}},\ and\
  \bibinfo {author} {\bibfnamefont {Y.}~\bibnamefont {Ohta}},\ }\bibfield
  {title} {\bibinfo {title} {Excitonic insulator state in the two-orbital
  hubbard model: Variational cluster approach},\ }\href@noop {} {\bibfield
  {journal} {\bibinfo  {journal} {Physical Review B—Condensed Matter and
  Materials Physics}\ }\textbf {\bibinfo {volume} {85}},\ \bibinfo {pages}
  {165135} (\bibinfo {year} {2012})}\BibitemShut {NoStop}%
\bibitem [{\citenamefont {Seki}\ \emph {et~al.}(2014)\citenamefont {Seki},
  \citenamefont {Wakisaka}, \citenamefont {Kaneko}, \citenamefont {Toriyama},
  \citenamefont {Konishi}, \citenamefont {Sudayama}, \citenamefont {Saini},
  \citenamefont {Arita}, \citenamefont {Namatame}, \citenamefont {Taniguchi}
  \emph {et~al.}}]{seki2014excitonic}%
  \BibitemOpen
  \bibfield  {author} {\bibinfo {author} {\bibfnamefont {K.}~\bibnamefont
  {Seki}}, \bibinfo {author} {\bibfnamefont {Y.}~\bibnamefont {Wakisaka}},
  \bibinfo {author} {\bibfnamefont {T.}~\bibnamefont {Kaneko}}, \bibinfo
  {author} {\bibfnamefont {T.}~\bibnamefont {Toriyama}}, \bibinfo {author}
  {\bibfnamefont {T.}~\bibnamefont {Konishi}}, \bibinfo {author} {\bibfnamefont
  {T.}~\bibnamefont {Sudayama}}, \bibinfo {author} {\bibfnamefont {N.~L.}\
  \bibnamefont {Saini}}, \bibinfo {author} {\bibfnamefont {M.}~\bibnamefont
  {Arita}}, \bibinfo {author} {\bibfnamefont {H.}~\bibnamefont {Namatame}},
  \bibinfo {author} {\bibfnamefont {M.}~\bibnamefont {Taniguchi}}, \emph
  {et~al.},\ }\bibfield  {title} {\bibinfo {title} {Excitonic bose-einstein
  condensation in ta 2 nise 5 above room temperature},\ }\href@noop {}
  {\bibfield  {journal} {\bibinfo  {journal} {Physical Review B}\ }\textbf
  {\bibinfo {volume} {90}},\ \bibinfo {pages} {155116} (\bibinfo {year}
  {2014})}\BibitemShut {NoStop}%
\bibitem [{\citenamefont {Monney}\ \emph {et~al.}(2009)\citenamefont {Monney},
  \citenamefont {Cercellier}, \citenamefont {Clerc}, \citenamefont {Battaglia},
  \citenamefont {Schwier}, \citenamefont {Didiot}, \citenamefont {Garnier},
  \citenamefont {Beck}, \citenamefont {Aebi}, \citenamefont {Berger} \emph
  {et~al.}}]{monney2009spontaneous}%
  \BibitemOpen
  \bibfield  {author} {\bibinfo {author} {\bibfnamefont {C.}~\bibnamefont
  {Monney}}, \bibinfo {author} {\bibfnamefont {H.}~\bibnamefont {Cercellier}},
  \bibinfo {author} {\bibfnamefont {F.}~\bibnamefont {Clerc}}, \bibinfo
  {author} {\bibfnamefont {C.}~\bibnamefont {Battaglia}}, \bibinfo {author}
  {\bibfnamefont {E.}~\bibnamefont {Schwier}}, \bibinfo {author} {\bibfnamefont
  {C.}~\bibnamefont {Didiot}}, \bibinfo {author} {\bibfnamefont {M.~G.}\
  \bibnamefont {Garnier}}, \bibinfo {author} {\bibfnamefont {H.}~\bibnamefont
  {Beck}}, \bibinfo {author} {\bibfnamefont {P.}~\bibnamefont {Aebi}}, \bibinfo
  {author} {\bibfnamefont {H.}~\bibnamefont {Berger}}, \emph {et~al.},\
  }\bibfield  {title} {\bibinfo {title} {Spontaneous exciton condensation in 1
  t-tise 2: Bcs-like approach},\ }\href@noop {} {\bibfield  {journal} {\bibinfo
   {journal} {Physical Review B—Condensed Matter and Materials Physics}\
  }\textbf {\bibinfo {volume} {79}},\ \bibinfo {pages} {045116} (\bibinfo
  {year} {2009})}\BibitemShut {NoStop}%
\bibitem [{\citenamefont {Monney}\ \emph {et~al.}(2012)\citenamefont {Monney},
  \citenamefont {Monney}, \citenamefont {Aebi},\ and\ \citenamefont
  {Beck}}]{monney2012electron}%
  \BibitemOpen
  \bibfield  {author} {\bibinfo {author} {\bibfnamefont {C.}~\bibnamefont
  {Monney}}, \bibinfo {author} {\bibfnamefont {G.}~\bibnamefont {Monney}},
  \bibinfo {author} {\bibfnamefont {P.}~\bibnamefont {Aebi}},\ and\ \bibinfo
  {author} {\bibfnamefont {H.}~\bibnamefont {Beck}},\ }\bibfield  {title}
  {\bibinfo {title} {Electron--hole instability in 1t-tise2},\ }\href@noop {}
  {\bibfield  {journal} {\bibinfo  {journal} {New Journal of Physics}\ }\textbf
  {\bibinfo {volume} {14}},\ \bibinfo {pages} {075026} (\bibinfo {year}
  {2012})}\BibitemShut {NoStop}%
\bibitem [{\citenamefont {Kaneko}\ \emph {et~al.}(2018)\citenamefont {Kaneko},
  \citenamefont {Ohta},\ and\ \citenamefont {Yunoki}}]{kaneko2018exciton}%
  \BibitemOpen
  \bibfield  {author} {\bibinfo {author} {\bibfnamefont {T.}~\bibnamefont
  {Kaneko}}, \bibinfo {author} {\bibfnamefont {Y.}~\bibnamefont {Ohta}},\ and\
  \bibinfo {author} {\bibfnamefont {S.}~\bibnamefont {Yunoki}},\ }\bibfield
  {title} {\bibinfo {title} {Exciton-phonon cooperative mechanism of the
  triple-q charge-density-wave and antiferroelectric electron polarization in
  tise 2},\ }\href@noop {} {\bibfield  {journal} {\bibinfo  {journal} {Physical
  Review B}\ }\textbf {\bibinfo {volume} {97}},\ \bibinfo {pages} {155131}
  (\bibinfo {year} {2018})}\BibitemShut {NoStop}%
\bibitem [{\citenamefont {Mor}\ \emph {et~al.}(2017)\citenamefont {Mor},
  \citenamefont {Herzog}, \citenamefont {Gole{\v{z}}}, \citenamefont {Werner},
  \citenamefont {Eckstein}, \citenamefont {Katayama}, \citenamefont {Nohara},
  \citenamefont {Takagi}, \citenamefont {Mizokawa}, \citenamefont {Monney}
  \emph {et~al.}}]{mor2017ultrafast}%
  \BibitemOpen
  \bibfield  {author} {\bibinfo {author} {\bibfnamefont {S.}~\bibnamefont
  {Mor}}, \bibinfo {author} {\bibfnamefont {M.}~\bibnamefont {Herzog}},
  \bibinfo {author} {\bibfnamefont {D.}~\bibnamefont {Gole{\v{z}}}}, \bibinfo
  {author} {\bibfnamefont {P.}~\bibnamefont {Werner}}, \bibinfo {author}
  {\bibfnamefont {M.}~\bibnamefont {Eckstein}}, \bibinfo {author}
  {\bibfnamefont {N.}~\bibnamefont {Katayama}}, \bibinfo {author}
  {\bibfnamefont {M.}~\bibnamefont {Nohara}}, \bibinfo {author} {\bibfnamefont
  {H.}~\bibnamefont {Takagi}}, \bibinfo {author} {\bibfnamefont
  {T.}~\bibnamefont {Mizokawa}}, \bibinfo {author} {\bibfnamefont
  {C.}~\bibnamefont {Monney}}, \emph {et~al.},\ }\bibfield  {title} {\bibinfo
  {title} {Ultrafast electronic band gap control in an excitonic insulator},\
  }\href@noop {} {\bibfield  {journal} {\bibinfo  {journal} {Physical review
  letters}\ }\textbf {\bibinfo {volume} {119}},\ \bibinfo {pages} {086401}
  (\bibinfo {year} {2017})}\BibitemShut {NoStop}%
\bibitem [{\citenamefont {Okazaki}\ \emph {et~al.}(2018)\citenamefont
  {Okazaki}, \citenamefont {Ogawa}, \citenamefont {Suzuki}, \citenamefont
  {Yamamoto}, \citenamefont {Someya}, \citenamefont {Michimae}, \citenamefont
  {Watanabe}, \citenamefont {Lu}, \citenamefont {Nohara}, \citenamefont
  {Takagi} \emph {et~al.}}]{okazaki2018photo}%
  \BibitemOpen
  \bibfield  {author} {\bibinfo {author} {\bibfnamefont {K.}~\bibnamefont
  {Okazaki}}, \bibinfo {author} {\bibfnamefont {Y.}~\bibnamefont {Ogawa}},
  \bibinfo {author} {\bibfnamefont {T.}~\bibnamefont {Suzuki}}, \bibinfo
  {author} {\bibfnamefont {T.}~\bibnamefont {Yamamoto}}, \bibinfo {author}
  {\bibfnamefont {T.}~\bibnamefont {Someya}}, \bibinfo {author} {\bibfnamefont
  {S.}~\bibnamefont {Michimae}}, \bibinfo {author} {\bibfnamefont
  {M.}~\bibnamefont {Watanabe}}, \bibinfo {author} {\bibfnamefont
  {Y.}~\bibnamefont {Lu}}, \bibinfo {author} {\bibfnamefont {M.}~\bibnamefont
  {Nohara}}, \bibinfo {author} {\bibfnamefont {H.}~\bibnamefont {Takagi}},
  \emph {et~al.},\ }\bibfield  {title} {\bibinfo {title} {Photo-induced
  semimetallic states realised in electron--hole coupled insulators},\
  }\href@noop {} {\bibfield  {journal} {\bibinfo  {journal} {Nature
  communications}\ }\textbf {\bibinfo {volume} {9}},\ \bibinfo {pages} {4322}
  (\bibinfo {year} {2018})}\BibitemShut {NoStop}%
\bibitem [{\citenamefont {Tang}\ \emph {et~al.}(2020)\citenamefont {Tang},
  \citenamefont {Wang}, \citenamefont {Duan}, \citenamefont {Yang},
  \citenamefont {Huang}, \citenamefont {Guo}, \citenamefont {Qian},\ and\
  \citenamefont {Zhang}}]{tang2020non}%
  \BibitemOpen
  \bibfield  {author} {\bibinfo {author} {\bibfnamefont {T.}~\bibnamefont
  {Tang}}, \bibinfo {author} {\bibfnamefont {H.}~\bibnamefont {Wang}}, \bibinfo
  {author} {\bibfnamefont {S.}~\bibnamefont {Duan}}, \bibinfo {author}
  {\bibfnamefont {Y.}~\bibnamefont {Yang}}, \bibinfo {author} {\bibfnamefont
  {C.}~\bibnamefont {Huang}}, \bibinfo {author} {\bibfnamefont
  {Y.}~\bibnamefont {Guo}}, \bibinfo {author} {\bibfnamefont {D.}~\bibnamefont
  {Qian}},\ and\ \bibinfo {author} {\bibfnamefont {W.}~\bibnamefont {Zhang}},\
  }\bibfield  {title} {\bibinfo {title} {Non-coulomb strong electron-hole
  binding in ta 2 nise 5 revealed by time-and angle-resolved photoemission
  spectroscopy},\ }\href@noop {} {\bibfield  {journal} {\bibinfo  {journal}
  {Physical Review B}\ }\textbf {\bibinfo {volume} {101}},\ \bibinfo {pages}
  {235148} (\bibinfo {year} {2020})}\BibitemShut {NoStop}%
\bibitem [{\citenamefont {Baldini}\ \emph {et~al.}(2023)\citenamefont
  {Baldini}, \citenamefont {Zong}, \citenamefont {Choi}, \citenamefont {Lee},
  \citenamefont {Michael}, \citenamefont {Windgaetter}, \citenamefont {Mazin},
  \citenamefont {Latini}, \citenamefont {Azoury}, \citenamefont {Lv} \emph
  {et~al.}}]{baldini2023spontaneous}%
  \BibitemOpen
  \bibfield  {author} {\bibinfo {author} {\bibfnamefont {E.}~\bibnamefont
  {Baldini}}, \bibinfo {author} {\bibfnamefont {A.}~\bibnamefont {Zong}},
  \bibinfo {author} {\bibfnamefont {D.}~\bibnamefont {Choi}}, \bibinfo {author}
  {\bibfnamefont {C.}~\bibnamefont {Lee}}, \bibinfo {author} {\bibfnamefont
  {M.~H.}\ \bibnamefont {Michael}}, \bibinfo {author} {\bibfnamefont
  {L.}~\bibnamefont {Windgaetter}}, \bibinfo {author} {\bibfnamefont {I.~I.}\
  \bibnamefont {Mazin}}, \bibinfo {author} {\bibfnamefont {S.}~\bibnamefont
  {Latini}}, \bibinfo {author} {\bibfnamefont {D.}~\bibnamefont {Azoury}},
  \bibinfo {author} {\bibfnamefont {B.}~\bibnamefont {Lv}}, \emph {et~al.},\
  }\bibfield  {title} {\bibinfo {title} {The spontaneous symmetry breaking in
  ta2nise5 is structural in nature},\ }\href@noop {} {\bibfield  {journal}
  {\bibinfo  {journal} {Proceedings of the National Academy of Sciences}\
  }\textbf {\bibinfo {volume} {120}},\ \bibinfo {pages} {e2221688120} (\bibinfo
  {year} {2023})}\BibitemShut {NoStop}%
\bibitem [{\citenamefont {Kim}\ \emph {et~al.}(2020)\citenamefont {Kim},
  \citenamefont {Schulz}, \citenamefont {Takayama}, \citenamefont {Isobe},
  \citenamefont {Takagi},\ and\ \citenamefont {Kaiser}}]{kim2020phononic}%
  \BibitemOpen
  \bibfield  {author} {\bibinfo {author} {\bibfnamefont {M.-J.}\ \bibnamefont
  {Kim}}, \bibinfo {author} {\bibfnamefont {A.}~\bibnamefont {Schulz}},
  \bibinfo {author} {\bibfnamefont {T.}~\bibnamefont {Takayama}}, \bibinfo
  {author} {\bibfnamefont {M.}~\bibnamefont {Isobe}}, \bibinfo {author}
  {\bibfnamefont {H.}~\bibnamefont {Takagi}},\ and\ \bibinfo {author}
  {\bibfnamefont {S.}~\bibnamefont {Kaiser}},\ }\bibfield  {title} {\bibinfo
  {title} {Phononic soft mode and strong electronic background behavior across
  the structural phase transition in the excitonic insulator ta $ \_2 $ nise $
  \_5 $(with erratum)},\ }\href@noop {} {\bibfield  {journal} {\bibinfo
  {journal} {arXiv preprint arXiv:2007.01723}\ } (\bibinfo {year}
  {2020})}\BibitemShut {NoStop}%
\bibitem [{\citenamefont {Bretscher}\ \emph
  {et~al.}(2021{\natexlab{a}})\citenamefont {Bretscher}, \citenamefont
  {Andrich}, \citenamefont {Telang}, \citenamefont {Singh}, \citenamefont
  {Harnagea}, \citenamefont {Sood},\ and\ \citenamefont
  {Rao}}]{bretscher2021ultrafast}%
  \BibitemOpen
  \bibfield  {author} {\bibinfo {author} {\bibfnamefont {H.~M.}\ \bibnamefont
  {Bretscher}}, \bibinfo {author} {\bibfnamefont {P.}~\bibnamefont {Andrich}},
  \bibinfo {author} {\bibfnamefont {P.}~\bibnamefont {Telang}}, \bibinfo
  {author} {\bibfnamefont {A.}~\bibnamefont {Singh}}, \bibinfo {author}
  {\bibfnamefont {L.}~\bibnamefont {Harnagea}}, \bibinfo {author}
  {\bibfnamefont {A.~K.}\ \bibnamefont {Sood}},\ and\ \bibinfo {author}
  {\bibfnamefont {A.}~\bibnamefont {Rao}},\ }\bibfield  {title} {\bibinfo
  {title} {Ultrafast melting and recovery of collective order in the excitonic
  insulator ta2nise5},\ }\href@noop {} {\bibfield  {journal} {\bibinfo
  {journal} {Nature communications}\ }\textbf {\bibinfo {volume} {12}},\
  \bibinfo {pages} {1699} (\bibinfo {year} {2021}{\natexlab{a}})}\BibitemShut
  {NoStop}%
\bibitem [{\citenamefont {Davari}\ \emph {et~al.}(2024)\citenamefont {Davari},
  \citenamefont {Ataei},\ and\ \citenamefont {Kargarian}}]{davari2024optical}%
  \BibitemOpen
  \bibfield  {author} {\bibinfo {author} {\bibfnamefont {E.}~\bibnamefont
  {Davari}}, \bibinfo {author} {\bibfnamefont {S.~S.}\ \bibnamefont {Ataei}},\
  and\ \bibinfo {author} {\bibfnamefont {M.}~\bibnamefont {Kargarian}},\
  }\bibfield  {title} {\bibinfo {title} {Optical drive of amplitude and phase
  modes in excitonic insulators},\ }\href@noop {} {\bibfield  {journal}
  {\bibinfo  {journal} {Physical Review B}\ }\textbf {\bibinfo {volume}
  {109}},\ \bibinfo {pages} {075146} (\bibinfo {year} {2024})}\BibitemShut
  {NoStop}%
\bibitem [{\citenamefont {Bretscher}\ \emph
  {et~al.}(2021{\natexlab{b}})\citenamefont {Bretscher}, \citenamefont
  {Andrich}, \citenamefont {Murakami}, \citenamefont {Gole{\v{z}}},
  \citenamefont {Remez}, \citenamefont {Telang}, \citenamefont {Singh},
  \citenamefont {Harnagea}, \citenamefont {Cooper}, \citenamefont {Millis}
  \emph {et~al.}}]{bretscher2021imaging}%
  \BibitemOpen
  \bibfield  {author} {\bibinfo {author} {\bibfnamefont {H.~M.}\ \bibnamefont
  {Bretscher}}, \bibinfo {author} {\bibfnamefont {P.}~\bibnamefont {Andrich}},
  \bibinfo {author} {\bibfnamefont {Y.}~\bibnamefont {Murakami}}, \bibinfo
  {author} {\bibfnamefont {D.}~\bibnamefont {Gole{\v{z}}}}, \bibinfo {author}
  {\bibfnamefont {B.}~\bibnamefont {Remez}}, \bibinfo {author} {\bibfnamefont
  {P.}~\bibnamefont {Telang}}, \bibinfo {author} {\bibfnamefont
  {A.}~\bibnamefont {Singh}}, \bibinfo {author} {\bibfnamefont
  {L.}~\bibnamefont {Harnagea}}, \bibinfo {author} {\bibfnamefont {N.~R.}\
  \bibnamefont {Cooper}}, \bibinfo {author} {\bibfnamefont {A.~J.}\
  \bibnamefont {Millis}}, \emph {et~al.},\ }\bibfield  {title} {\bibinfo
  {title} {Imaging the coherent propagation of collective modes in the
  excitonic insulator ta2nise5 at room temperature},\ }\href@noop {} {\bibfield
   {journal} {\bibinfo  {journal} {Science Advances}\ }\textbf {\bibinfo
  {volume} {7}},\ \bibinfo {pages} {eabd6147} (\bibinfo {year}
  {2021}{\natexlab{b}})}\BibitemShut {NoStop}%
\bibitem [{\citenamefont {Gole{\v{z}}}\ \emph {et~al.}(2022)\citenamefont
  {Gole{\v{z}}}, \citenamefont {Dufresne}, \citenamefont {Kim}, \citenamefont
  {Boschini}, \citenamefont {Chu}, \citenamefont {Murakami}, \citenamefont
  {Levy}, \citenamefont {Mills}, \citenamefont {Zhdanovich}, \citenamefont
  {Isobe} \emph {et~al.}}]{golevz2022unveiling}%
  \BibitemOpen
  \bibfield  {author} {\bibinfo {author} {\bibfnamefont {D.}~\bibnamefont
  {Gole{\v{z}}}}, \bibinfo {author} {\bibfnamefont {S.~K.}\ \bibnamefont
  {Dufresne}}, \bibinfo {author} {\bibfnamefont {M.-J.}\ \bibnamefont {Kim}},
  \bibinfo {author} {\bibfnamefont {F.}~\bibnamefont {Boschini}}, \bibinfo
  {author} {\bibfnamefont {H.}~\bibnamefont {Chu}}, \bibinfo {author}
  {\bibfnamefont {Y.}~\bibnamefont {Murakami}}, \bibinfo {author}
  {\bibfnamefont {G.}~\bibnamefont {Levy}}, \bibinfo {author} {\bibfnamefont
  {A.~K.}\ \bibnamefont {Mills}}, \bibinfo {author} {\bibfnamefont
  {S.}~\bibnamefont {Zhdanovich}}, \bibinfo {author} {\bibfnamefont
  {M.}~\bibnamefont {Isobe}}, \emph {et~al.},\ }\bibfield  {title} {\bibinfo
  {title} {Unveiling the underlying interactions in ta 2 nise 5 from
  photoinduced lifetime change},\ }\href@noop {} {\bibfield  {journal}
  {\bibinfo  {journal} {Physical Review B}\ }\textbf {\bibinfo {volume}
  {106}},\ \bibinfo {pages} {L121106} (\bibinfo {year} {2022})}\BibitemShut
  {NoStop}%
\bibitem [{\citenamefont {Sentef}\ \emph {et~al.}(2020)\citenamefont {Sentef},
  \citenamefont {Li}, \citenamefont {K{\"u}nzel},\ and\ \citenamefont
  {Eckstein}}]{sentef2020quantum}%
  \BibitemOpen
  \bibfield  {author} {\bibinfo {author} {\bibfnamefont {M.~A.}\ \bibnamefont
  {Sentef}}, \bibinfo {author} {\bibfnamefont {J.}~\bibnamefont {Li}}, \bibinfo
  {author} {\bibfnamefont {F.}~\bibnamefont {K{\"u}nzel}},\ and\ \bibinfo
  {author} {\bibfnamefont {M.}~\bibnamefont {Eckstein}},\ }\bibfield  {title}
  {\bibinfo {title} {Quantum to classical crossover of floquet engineering in
  correlated quantum systems},\ }\href@noop {} {\bibfield  {journal} {\bibinfo
  {journal} {Physical Review Research}\ }\textbf {\bibinfo {volume} {2}},\
  \bibinfo {pages} {033033} (\bibinfo {year} {2020})}\BibitemShut {NoStop}%
\bibitem [{\citenamefont {Ruggenthaler}\ \emph {et~al.}(2018)\citenamefont
  {Ruggenthaler}, \citenamefont {Tancogne-Dejean}, \citenamefont {Flick},
  \citenamefont {Appel},\ and\ \citenamefont
  {Rubio}}]{ruggenthaler2018quantum}%
  \BibitemOpen
  \bibfield  {author} {\bibinfo {author} {\bibfnamefont {M.}~\bibnamefont
  {Ruggenthaler}}, \bibinfo {author} {\bibfnamefont {N.}~\bibnamefont
  {Tancogne-Dejean}}, \bibinfo {author} {\bibfnamefont {J.}~\bibnamefont
  {Flick}}, \bibinfo {author} {\bibfnamefont {H.}~\bibnamefont {Appel}},\ and\
  \bibinfo {author} {\bibfnamefont {A.}~\bibnamefont {Rubio}},\ }\bibfield
  {title} {\bibinfo {title} {From a quantum-electrodynamical light--matter
  description to novel spectroscopies},\ }\href@noop {} {\bibfield  {journal}
  {\bibinfo  {journal} {Nature Reviews Chemistry}\ }\textbf {\bibinfo {volume}
  {2}},\ \bibinfo {pages} {1} (\bibinfo {year} {2018})}\BibitemShut {NoStop}%
\bibitem [{\citenamefont {Lysne}\ \emph {et~al.}(2023)\citenamefont {Lysne},
  \citenamefont {Sch{\"u}ler},\ and\ \citenamefont
  {Werner}}]{lysne2023quantum}%
  \BibitemOpen
  \bibfield  {author} {\bibinfo {author} {\bibfnamefont {M.}~\bibnamefont
  {Lysne}}, \bibinfo {author} {\bibfnamefont {M.}~\bibnamefont {Sch{\"u}ler}},\
  and\ \bibinfo {author} {\bibfnamefont {P.}~\bibnamefont {Werner}},\
  }\bibfield  {title} {\bibinfo {title} {Quantum optics measurement scheme for
  quantum geometry and topological invariants},\ }\href@noop {} {\bibfield
  {journal} {\bibinfo  {journal} {Physical Review Letters}\ }\textbf {\bibinfo
  {volume} {131}},\ \bibinfo {pages} {156901} (\bibinfo {year}
  {2023})}\BibitemShut {NoStop}%
\bibitem [{\citenamefont {Dmytruk}\ and\ \citenamefont
  {Schir{\`o}}(2022)}]{dmytruk2022controlling}%
  \BibitemOpen
  \bibfield  {author} {\bibinfo {author} {\bibfnamefont {O.}~\bibnamefont
  {Dmytruk}}\ and\ \bibinfo {author} {\bibfnamefont {M.}~\bibnamefont
  {Schir{\`o}}},\ }\bibfield  {title} {\bibinfo {title} {Controlling
  topological phases of matter with quantum light},\ }\href@noop {} {\bibfield
  {journal} {\bibinfo  {journal} {Communications Physics}\ }\textbf {\bibinfo
  {volume} {5}},\ \bibinfo {pages} {271} (\bibinfo {year} {2022})}\BibitemShut
  {NoStop}%
\bibitem [{\citenamefont {Passetti}\ \emph {et~al.}(2023)\citenamefont
  {Passetti}, \citenamefont {Eckhardt}, \citenamefont {Sentef},\ and\
  \citenamefont {Kennes}}]{passetti2023cavity}%
  \BibitemOpen
  \bibfield  {author} {\bibinfo {author} {\bibfnamefont {G.}~\bibnamefont
  {Passetti}}, \bibinfo {author} {\bibfnamefont {C.~J.}\ \bibnamefont
  {Eckhardt}}, \bibinfo {author} {\bibfnamefont {M.~A.}\ \bibnamefont
  {Sentef}},\ and\ \bibinfo {author} {\bibfnamefont {D.~M.}\ \bibnamefont
  {Kennes}},\ }\bibfield  {title} {\bibinfo {title} {Cavity light-matter
  entanglement through quantum fluctuations},\ }\href@noop {} {\bibfield
  {journal} {\bibinfo  {journal} {Physical Review Letters}\ }\textbf {\bibinfo
  {volume} {131}},\ \bibinfo {pages} {023601} (\bibinfo {year}
  {2023})}\BibitemShut {NoStop}%
\bibitem [{\citenamefont {Lenk}\ and\ \citenamefont
  {Eckstein}(2020)}]{lenk2020collective}%
  \BibitemOpen
  \bibfield  {author} {\bibinfo {author} {\bibfnamefont {K.}~\bibnamefont
  {Lenk}}\ and\ \bibinfo {author} {\bibfnamefont {M.}~\bibnamefont
  {Eckstein}},\ }\bibfield  {title} {\bibinfo {title} {Collective excitations
  of the u (1)-symmetric exciton insulator in a cavity},\ }\href@noop {}
  {\bibfield  {journal} {\bibinfo  {journal} {Physical Review B}\ }\textbf
  {\bibinfo {volume} {102}},\ \bibinfo {pages} {205129} (\bibinfo {year}
  {2020})}\BibitemShut {NoStop}%
\bibitem [{\citenamefont {Frisk~Kockum}\ \emph {et~al.}(2019)\citenamefont
  {Frisk~Kockum}, \citenamefont {Miranowicz}, \citenamefont {De~Liberato},
  \citenamefont {Savasta},\ and\ \citenamefont {Nori}}]{frisk2019ultrastrong}%
  \BibitemOpen
  \bibfield  {author} {\bibinfo {author} {\bibfnamefont {A.}~\bibnamefont
  {Frisk~Kockum}}, \bibinfo {author} {\bibfnamefont {A.}~\bibnamefont
  {Miranowicz}}, \bibinfo {author} {\bibfnamefont {S.}~\bibnamefont
  {De~Liberato}}, \bibinfo {author} {\bibfnamefont {S.}~\bibnamefont
  {Savasta}},\ and\ \bibinfo {author} {\bibfnamefont {F.}~\bibnamefont
  {Nori}},\ }\bibfield  {title} {\bibinfo {title} {Ultrastrong coupling between
  light and matter},\ }\href@noop {} {\bibfield  {journal} {\bibinfo  {journal}
  {Nature Reviews Physics}\ }\textbf {\bibinfo {volume} {1}},\ \bibinfo {pages}
  {19} (\bibinfo {year} {2019})}\BibitemShut {NoStop}%
\bibitem [{\citenamefont {Schlawin}\ \emph {et~al.}(2022)\citenamefont
  {Schlawin}, \citenamefont {Kennes},\ and\ \citenamefont
  {Sentef}}]{schlawin2022cavity}%
  \BibitemOpen
  \bibfield  {author} {\bibinfo {author} {\bibfnamefont {F.}~\bibnamefont
  {Schlawin}}, \bibinfo {author} {\bibfnamefont {D.~M.}\ \bibnamefont
  {Kennes}},\ and\ \bibinfo {author} {\bibfnamefont {M.~A.}\ \bibnamefont
  {Sentef}},\ }\bibfield  {title} {\bibinfo {title} {Cavity quantum
  materials},\ }\href@noop {} {\bibfield  {journal} {\bibinfo  {journal}
  {Applied Physics Reviews}\ }\textbf {\bibinfo {volume} {9}} (\bibinfo {year}
  {2022})}\BibitemShut {NoStop}%
\bibitem [{\citenamefont {Mazza}\ and\ \citenamefont
  {Georges}(2019)}]{mazza2019superradiant}%
  \BibitemOpen
  \bibfield  {author} {\bibinfo {author} {\bibfnamefont {G.}~\bibnamefont
  {Mazza}}\ and\ \bibinfo {author} {\bibfnamefont {A.}~\bibnamefont
  {Georges}},\ }\bibfield  {title} {\bibinfo {title} {Superradiant quantum
  materials},\ }\href@noop {} {\bibfield  {journal} {\bibinfo  {journal}
  {Physical review letters}\ }\textbf {\bibinfo {volume} {122}},\ \bibinfo
  {pages} {017401} (\bibinfo {year} {2019})}\BibitemShut {NoStop}%
\bibitem [{\citenamefont {Andolina}\ \emph {et~al.}(2019)\citenamefont
  {Andolina}, \citenamefont {Pellegrino}, \citenamefont {Giovannetti},
  \citenamefont {MacDonald},\ and\ \citenamefont
  {Polini}}]{andolina2019cavity}%
  \BibitemOpen
  \bibfield  {author} {\bibinfo {author} {\bibfnamefont {G.}~\bibnamefont
  {Andolina}}, \bibinfo {author} {\bibfnamefont {F.}~\bibnamefont
  {Pellegrino}}, \bibinfo {author} {\bibfnamefont {V.}~\bibnamefont
  {Giovannetti}}, \bibinfo {author} {\bibfnamefont {A.}~\bibnamefont
  {MacDonald}},\ and\ \bibinfo {author} {\bibfnamefont {M.}~\bibnamefont
  {Polini}},\ }\bibfield  {title} {\bibinfo {title} {Cavity quantum
  electrodynamics of strongly correlated electron systems: A no-go theorem for
  photon condensation},\ }\href@noop {} {\bibfield  {journal} {\bibinfo
  {journal} {Physical Review B}\ }\textbf {\bibinfo {volume} {100}},\ \bibinfo
  {pages} {121109} (\bibinfo {year} {2019})}\BibitemShut {NoStop}%
\bibitem [{\citenamefont {Shockley}(1939)}]{shockley1939surface}%
  \BibitemOpen
  \bibfield  {author} {\bibinfo {author} {\bibfnamefont {W.}~\bibnamefont
  {Shockley}},\ }\bibfield  {title} {\bibinfo {title} {On the surface states
  associated with a periodic potential},\ }\href@noop {} {\bibfield  {journal}
  {\bibinfo  {journal} {Physical review}\ }\textbf {\bibinfo {volume} {56}},\
  \bibinfo {pages} {317} (\bibinfo {year} {1939})}\BibitemShut {NoStop}%
\bibitem [{\citenamefont {Khatibi}\ \emph {et~al.}(2020)\citenamefont
  {Khatibi}, \citenamefont {Ahemeh},\ and\ \citenamefont
  {Kargarian}}]{khatibi2020excitonic}%
  \BibitemOpen
  \bibfield  {author} {\bibinfo {author} {\bibfnamefont {Z.}~\bibnamefont
  {Khatibi}}, \bibinfo {author} {\bibfnamefont {R.}~\bibnamefont {Ahemeh}},\
  and\ \bibinfo {author} {\bibfnamefont {M.}~\bibnamefont {Kargarian}},\
  }\bibfield  {title} {\bibinfo {title} {Excitonic insulator phase and
  condensate dynamics in a topological one-dimensional model},\ }\href@noop {}
  {\bibfield  {journal} {\bibinfo  {journal} {Physical Review B}\ }\textbf
  {\bibinfo {volume} {102}},\ \bibinfo {pages} {245121} (\bibinfo {year}
  {2020})}\BibitemShut {NoStop}%
\bibitem [{\citenamefont {Kune{\v{s}}}(2015)}]{kunevs2015excitonic}%
  \BibitemOpen
  \bibfield  {author} {\bibinfo {author} {\bibfnamefont {J.}~\bibnamefont
  {Kune{\v{s}}}},\ }\bibfield  {title} {\bibinfo {title} {Excitonic
  condensation in systems of strongly correlated electrons},\ }\href@noop {}
  {\bibfield  {journal} {\bibinfo  {journal} {Journal of Physics: Condensed
  Matter}\ }\textbf {\bibinfo {volume} {27}},\ \bibinfo {pages} {333201}
  (\bibinfo {year} {2015})}\BibitemShut {NoStop}%
\bibitem [{\citenamefont {Vogel}\ and\ \citenamefont
  {Welsch}(2006)}]{vogel2006quantum}%
  \BibitemOpen
  \bibfield  {author} {\bibinfo {author} {\bibfnamefont {W.}~\bibnamefont
  {Vogel}}\ and\ \bibinfo {author} {\bibfnamefont {D.-G.}\ \bibnamefont
  {Welsch}},\ }\href@noop {} {\emph {\bibinfo {title} {Quantum optics}}}\
  (\bibinfo  {publisher} {John Wiley \& Sons},\ \bibinfo {year}
  {2006})\BibitemShut {NoStop}%
\bibitem [{\citenamefont {Loudon}(2000)}]{loudon2000quantum}%
  \BibitemOpen
  \bibfield  {author} {\bibinfo {author} {\bibfnamefont {R.}~\bibnamefont
  {Loudon}},\ }\href@noop {} {\emph {\bibinfo {title} {The quantum theory of
  light}}}\ (\bibinfo  {publisher} {OUP Oxford},\ \bibinfo {year}
  {2000})\BibitemShut {NoStop}%
\bibitem [{\citenamefont {Grynberg}\ \emph {et~al.}(2010)\citenamefont
  {Grynberg}, \citenamefont {Aspect},\ and\ \citenamefont
  {Fabre}}]{grynberg2010introduction}%
  \BibitemOpen
  \bibfield  {author} {\bibinfo {author} {\bibfnamefont {G.}~\bibnamefont
  {Grynberg}}, \bibinfo {author} {\bibfnamefont {A.}~\bibnamefont {Aspect}},\
  and\ \bibinfo {author} {\bibfnamefont {C.}~\bibnamefont {Fabre}},\
  }\href@noop {} {\emph {\bibinfo {title} {Introduction to quantum optics: from
  the semi-classical approach to quantized light}}}\ (\bibinfo  {publisher}
  {Cambridge university press},\ \bibinfo {year} {2010})\BibitemShut {NoStop}%
\bibitem [{\citenamefont {Davari}\ and\ \citenamefont
  {Kargarian}(2025)}]{DataAvailiblity_CavityEI}%
  \BibitemOpen
  \bibfield  {author} {\bibinfo {author} {\bibfnamefont {E.}~\bibnamefont
  {Davari}}\ and\ \bibinfo {author} {\bibfnamefont {M.}~\bibnamefont
  {Kargarian}},\ }\bibfield  {title} {\bibinfo {title} {Data availability: The
  matlab codes are vailable in the published page as supplementary materials},\
  }\href@noop {} {\  (\bibinfo {year} {2025})}\BibitemShut {NoStop}%
\bibitem [{\citenamefont {Anderson}(1958)}]{anderson1958random}%
  \BibitemOpen
  \bibfield  {author} {\bibinfo {author} {\bibfnamefont {P.~W.}\ \bibnamefont
  {Anderson}},\ }\bibfield  {title} {\bibinfo {title} {Random-phase
  approximation in the theory of superconductivity},\ }\href@noop {} {\bibfield
   {journal} {\bibinfo  {journal} {Physical Review}\ }\textbf {\bibinfo
  {volume} {112}},\ \bibinfo {pages} {1900} (\bibinfo {year}
  {1958})}\BibitemShut {NoStop}%
\bibitem [{\citenamefont {Gole{\v{z}}}\ \emph {et~al.}(2020)\citenamefont
  {Gole{\v{z}}}, \citenamefont {Sun}, \citenamefont {Murakami}, \citenamefont
  {Georges},\ and\ \citenamefont {Millis}}]{golevz2020nonlinear}%
  \BibitemOpen
  \bibfield  {author} {\bibinfo {author} {\bibfnamefont {D.}~\bibnamefont
  {Gole{\v{z}}}}, \bibinfo {author} {\bibfnamefont {Z.}~\bibnamefont {Sun}},
  \bibinfo {author} {\bibfnamefont {Y.}~\bibnamefont {Murakami}}, \bibinfo
  {author} {\bibfnamefont {A.}~\bibnamefont {Georges}},\ and\ \bibinfo {author}
  {\bibfnamefont {A.~J.}\ \bibnamefont {Millis}},\ }\bibfield  {title}
  {\bibinfo {title} {Nonlinear spectroscopy of collective modes in an excitonic
  insulator},\ }\href@noop {} {\bibfield  {journal} {\bibinfo  {journal}
  {Physical Review Letters}\ }\textbf {\bibinfo {volume} {125}},\ \bibinfo
  {pages} {257601} (\bibinfo {year} {2020})}\BibitemShut {NoStop}%
\bibitem [{\citenamefont {Murakami}\ \emph {et~al.}(2017)\citenamefont
  {Murakami}, \citenamefont {Gole{\v{z}}}, \citenamefont {Eckstein},\ and\
  \citenamefont {Werner}}]{murakami2017photoinduced}%
  \BibitemOpen
  \bibfield  {author} {\bibinfo {author} {\bibfnamefont {Y.}~\bibnamefont
  {Murakami}}, \bibinfo {author} {\bibfnamefont {D.}~\bibnamefont
  {Gole{\v{z}}}}, \bibinfo {author} {\bibfnamefont {M.}~\bibnamefont
  {Eckstein}},\ and\ \bibinfo {author} {\bibfnamefont {P.}~\bibnamefont
  {Werner}},\ }\bibfield  {title} {\bibinfo {title} {Photoinduced enhancement
  of excitonic order},\ }\href@noop {} {\bibfield  {journal} {\bibinfo
  {journal} {Physical review letters}\ }\textbf {\bibinfo {volume} {119}},\
  \bibinfo {pages} {247601} (\bibinfo {year} {2017})}\BibitemShut {NoStop}%
\bibitem [{\citenamefont {Murakami}\ \emph {et~al.}(2020)\citenamefont
  {Murakami}, \citenamefont {Gole{\v{z}}}, \citenamefont {Kaneko},
  \citenamefont {Koga}, \citenamefont {Millis},\ and\ \citenamefont
  {Werner}}]{murakami2020collective}%
  \BibitemOpen
  \bibfield  {author} {\bibinfo {author} {\bibfnamefont {Y.}~\bibnamefont
  {Murakami}}, \bibinfo {author} {\bibfnamefont {D.}~\bibnamefont
  {Gole{\v{z}}}}, \bibinfo {author} {\bibfnamefont {T.}~\bibnamefont {Kaneko}},
  \bibinfo {author} {\bibfnamefont {A.}~\bibnamefont {Koga}}, \bibinfo {author}
  {\bibfnamefont {A.~J.}\ \bibnamefont {Millis}},\ and\ \bibinfo {author}
  {\bibfnamefont {P.}~\bibnamefont {Werner}},\ }\bibfield  {title} {\bibinfo
  {title} {Collective modes in excitonic insulators: Effects of electron-phonon
  coupling and signatures in the optical response},\ }\href@noop {} {\bibfield
  {journal} {\bibinfo  {journal} {Physical Review B}\ }\textbf {\bibinfo
  {volume} {101}},\ \bibinfo {pages} {195118} (\bibinfo {year}
  {2020})}\BibitemShut {NoStop}%
\bibitem [{\citenamefont {Kaneko}\ \emph {et~al.}(2021)\citenamefont {Kaneko},
  \citenamefont {Sun}, \citenamefont {Murakami}, \citenamefont {Gole{\v{z}}},\
  and\ \citenamefont {Millis}}]{kaneko2021bulk}%
  \BibitemOpen
  \bibfield  {author} {\bibinfo {author} {\bibfnamefont {T.}~\bibnamefont
  {Kaneko}}, \bibinfo {author} {\bibfnamefont {Z.}~\bibnamefont {Sun}},
  \bibinfo {author} {\bibfnamefont {Y.}~\bibnamefont {Murakami}}, \bibinfo
  {author} {\bibfnamefont {D.}~\bibnamefont {Gole{\v{z}}}},\ and\ \bibinfo
  {author} {\bibfnamefont {A.~J.}\ \bibnamefont {Millis}},\ }\bibfield  {title}
  {\bibinfo {title} {Bulk photovoltaic effect driven by collective excitations
  in a correlated insulator},\ }\href@noop {} {\bibfield  {journal} {\bibinfo
  {journal} {Physical Review Letters}\ }\textbf {\bibinfo {volume} {127}},\
  \bibinfo {pages} {127402} (\bibinfo {year} {2021})}\BibitemShut {NoStop}%
\bibitem [{\citenamefont {Li}\ \emph {et~al.}(2020)\citenamefont {Li},
  \citenamefont {Golez}, \citenamefont {Mazza}, \citenamefont {Millis},
  \citenamefont {Georges},\ and\ \citenamefont
  {Eckstein}}]{li2020electromagnetic}%
  \BibitemOpen
  \bibfield  {author} {\bibinfo {author} {\bibfnamefont {J.}~\bibnamefont
  {Li}}, \bibinfo {author} {\bibfnamefont {D.}~\bibnamefont {Golez}}, \bibinfo
  {author} {\bibfnamefont {G.}~\bibnamefont {Mazza}}, \bibinfo {author}
  {\bibfnamefont {A.~J.}\ \bibnamefont {Millis}}, \bibinfo {author}
  {\bibfnamefont {A.}~\bibnamefont {Georges}},\ and\ \bibinfo {author}
  {\bibfnamefont {M.}~\bibnamefont {Eckstein}},\ }\bibfield  {title} {\bibinfo
  {title} {Electromagnetic coupling in tight-binding models for strongly
  correlated light and matter},\ }\href@noop {} {\bibfield  {journal} {\bibinfo
   {journal} {Physical Review B}\ }\textbf {\bibinfo {volume} {101}},\ \bibinfo
  {pages} {205140} (\bibinfo {year} {2020})}\BibitemShut {NoStop}%
\bibitem [{\citenamefont {Li}\ and\ \citenamefont
  {Eckstein}(2020)}]{li2020manipulating}%
  \BibitemOpen
  \bibfield  {author} {\bibinfo {author} {\bibfnamefont {J.}~\bibnamefont
  {Li}}\ and\ \bibinfo {author} {\bibfnamefont {M.}~\bibnamefont {Eckstein}},\
  }\bibfield  {title} {\bibinfo {title} {Manipulating intertwined orders in
  solids with quantum light},\ }\href@noop {} {\bibfield  {journal} {\bibinfo
  {journal} {Physical Review Letters}\ }\textbf {\bibinfo {volume} {125}},\
  \bibinfo {pages} {217402} (\bibinfo {year} {2020})}\BibitemShut {NoStop}%
\bibitem [{\citenamefont {Guerci}\ \emph {et~al.}(2020)\citenamefont {Guerci},
  \citenamefont {Simon},\ and\ \citenamefont {Mora}}]{guerci2020superradiant}%
  \BibitemOpen
  \bibfield  {author} {\bibinfo {author} {\bibfnamefont {D.}~\bibnamefont
  {Guerci}}, \bibinfo {author} {\bibfnamefont {P.}~\bibnamefont {Simon}},\ and\
  \bibinfo {author} {\bibfnamefont {C.}~\bibnamefont {Mora}},\ }\bibfield
  {title} {\bibinfo {title} {Superradiant phase transition in electronic
  systems and emergent topological phases},\ }\href@noop {} {\bibfield
  {journal} {\bibinfo  {journal} {Physical Review Letters}\ }\textbf {\bibinfo
  {volume} {125}},\ \bibinfo {pages} {257604} (\bibinfo {year}
  {2020})}\BibitemShut {NoStop}%
\bibitem [{\citenamefont {Dmytruk}\ and\ \citenamefont
  {Schir{\'o}}(2021)}]{dmytruk2021gauge}%
  \BibitemOpen
  \bibfield  {author} {\bibinfo {author} {\bibfnamefont {O.}~\bibnamefont
  {Dmytruk}}\ and\ \bibinfo {author} {\bibfnamefont {M.}~\bibnamefont
  {Schir{\'o}}},\ }\bibfield  {title} {\bibinfo {title} {Gauge fixing for
  strongly correlated electrons coupled to quantum light},\ }\href@noop {}
  {\bibfield  {journal} {\bibinfo  {journal} {Physical Review B}\ }\textbf
  {\bibinfo {volume} {103}},\ \bibinfo {pages} {075131} (\bibinfo {year}
  {2021})}\BibitemShut {NoStop}%
\bibitem [{\citenamefont {Viviescas}\ and\ \citenamefont
  {Hackenbroich}(2003)}]{viviescas2003field}%
  \BibitemOpen
  \bibfield  {author} {\bibinfo {author} {\bibfnamefont {C.}~\bibnamefont
  {Viviescas}}\ and\ \bibinfo {author} {\bibfnamefont {G.}~\bibnamefont
  {Hackenbroich}},\ }\bibfield  {title} {\bibinfo {title} {Field quantization
  for open optical cavities},\ }\href@noop {} {\bibfield  {journal} {\bibinfo
  {journal} {Physical Review A}\ }\textbf {\bibinfo {volume} {67}},\ \bibinfo
  {pages} {013805} (\bibinfo {year} {2003})}\BibitemShut {NoStop}%
\bibitem [{\citenamefont {Jiang}\ and\ \citenamefont
  {Luu}(2008)}]{jiang2008heterodyne}%
  \BibitemOpen
  \bibfield  {author} {\bibinfo {author} {\bibfnamefont {L.~A.}\ \bibnamefont
  {Jiang}}\ and\ \bibinfo {author} {\bibfnamefont {J.~X.}\ \bibnamefont
  {Luu}},\ }\bibfield  {title} {\bibinfo {title} {Heterodyne detection with a
  weak local oscillator},\ }\href@noop {} {\bibfield  {journal} {\bibinfo
  {journal} {Applied optics}\ }\textbf {\bibinfo {volume} {47}},\ \bibinfo
  {pages} {1486} (\bibinfo {year} {2008})}\BibitemShut {NoStop}%
\bibitem [{\citenamefont {Mazza}\ and\ \citenamefont
  {Polini}(2023)}]{Mazza2023}%
  \BibitemOpen
  \bibfield  {author} {\bibinfo {author} {\bibfnamefont {G.}~\bibnamefont
  {Mazza}}\ and\ \bibinfo {author} {\bibfnamefont {M.}~\bibnamefont {Polini}},\
  }\bibfield  {title} {\bibinfo {title} {Hidden excitonic quantum states with
  broken time-reversal symmetry},\ }\href
  {https://doi.org/10.1103/PhysRevB.108.L241107} {\bibfield  {journal}
  {\bibinfo  {journal} {Phys. Rev. B}\ }\textbf {\bibinfo {volume} {108}},\
  \bibinfo {pages} {L241107} (\bibinfo {year} {2023})}\BibitemShut {NoStop}%
\bibitem [{\citenamefont {Scully}\ and\ \citenamefont
  {Zubairy}(1997)}]{scully1997quantum}%
  \BibitemOpen
  \bibfield  {author} {\bibinfo {author} {\bibfnamefont {M.~O.}\ \bibnamefont
  {Scully}}\ and\ \bibinfo {author} {\bibfnamefont {M.~S.}\ \bibnamefont
  {Zubairy}},\ }\href@noop {} {\emph {\bibinfo {title} {Quantum optics}}}\
  (\bibinfo  {publisher} {Cambridge university press},\ \bibinfo {year}
  {1997})\BibitemShut {NoStop}%
\bibitem [{\citenamefont {Haroche}\ and\ \citenamefont
  {Kleppner}(1989)}]{haroche1989cavity}%
  \BibitemOpen
  \bibfield  {author} {\bibinfo {author} {\bibfnamefont {S.}~\bibnamefont
  {Haroche}}\ and\ \bibinfo {author} {\bibfnamefont {D.}~\bibnamefont
  {Kleppner}},\ }\bibfield  {title} {\bibinfo {title} {Cavity quantum
  electrodynamics},\ }\href@noop {} {\bibfield  {journal} {\bibinfo  {journal}
  {Physics Today}\ }\textbf {\bibinfo {volume} {42}},\ \bibinfo {pages} {24}
  (\bibinfo {year} {1989})}\BibitemShut {NoStop}%
\bibitem [{\citenamefont {Watanabe}\ \emph {et~al.}(2015)\citenamefont
  {Watanabe}, \citenamefont {Seki},\ and\ \citenamefont
  {Yunoki}}]{watanabe2015charge}%
  \BibitemOpen
  \bibfield  {author} {\bibinfo {author} {\bibfnamefont {H.}~\bibnamefont
  {Watanabe}}, \bibinfo {author} {\bibfnamefont {K.}~\bibnamefont {Seki}},\
  and\ \bibinfo {author} {\bibfnamefont {S.}~\bibnamefont {Yunoki}},\
  }\bibfield  {title} {\bibinfo {title} {Charge-density wave induced by
  combined electron-electron and electron-phonon interactions in 1 t-tise 2: A
  variational monte carlo study},\ }\href@noop {} {\bibfield  {journal}
  {\bibinfo  {journal} {Physical Review B}\ }\textbf {\bibinfo {volume} {91}},\
  \bibinfo {pages} {205135} (\bibinfo {year} {2015})}\BibitemShut {NoStop}%
\bibitem [{\citenamefont {Zenker}\ \emph {et~al.}(2013)\citenamefont {Zenker},
  \citenamefont {Fehske}, \citenamefont {Beck}, \citenamefont {Monney},\ and\
  \citenamefont {Bishop}}]{Zenker2013}%
  \BibitemOpen
  \bibfield  {author} {\bibinfo {author} {\bibfnamefont {B.}~\bibnamefont
  {Zenker}}, \bibinfo {author} {\bibfnamefont {H.}~\bibnamefont {Fehske}},
  \bibinfo {author} {\bibfnamefont {H.}~\bibnamefont {Beck}}, \bibinfo {author}
  {\bibfnamefont {C.}~\bibnamefont {Monney}},\ and\ \bibinfo {author}
  {\bibfnamefont {A.}~\bibnamefont {Bishop}},\ }\bibfield  {title} {\bibinfo
  {title} {Chiral charge order in 1 t-tise 2: Importance of lattice degrees of
  freedom},\ }\href@noop {} {\bibfield  {journal} {\bibinfo  {journal}
  {Physical Review B—Condensed Matter and Materials Physics}\ }\textbf
  {\bibinfo {volume} {88}},\ \bibinfo {pages} {075138} (\bibinfo {year}
  {2013})}\BibitemShut {NoStop}%
\bibitem [{\citenamefont {Phan}\ \emph {et~al.}(2014)\citenamefont {Phan},
  \citenamefont {Fehske},\ and\ \citenamefont {Becker}}]{phan2014linear}%
  \BibitemOpen
  \bibfield  {author} {\bibinfo {author} {\bibfnamefont {V.-N.}\ \bibnamefont
  {Phan}}, \bibinfo {author} {\bibfnamefont {H.}~\bibnamefont {Fehske}},\ and\
  \bibinfo {author} {\bibfnamefont {K.~W.}\ \bibnamefont {Becker}},\ }\bibfield
   {title} {\bibinfo {title} {Linear response within the projector-based
  renormalization method: many-body corrections beyond the random phase
  approximation},\ }\href@noop {} {\bibfield  {journal} {\bibinfo  {journal}
  {The European Physical Journal B}\ }\textbf {\bibinfo {volume} {87}},\
  \bibinfo {pages} {1} (\bibinfo {year} {2014})}\BibitemShut {NoStop}%
\bibitem [{\citenamefont {Zenker}\ \emph {et~al.}(2014)\citenamefont {Zenker},
  \citenamefont {Fehske},\ and\ \citenamefont {Beck}}]{zenker2014fate}%
  \BibitemOpen
  \bibfield  {author} {\bibinfo {author} {\bibfnamefont {B.}~\bibnamefont
  {Zenker}}, \bibinfo {author} {\bibfnamefont {H.}~\bibnamefont {Fehske}},\
  and\ \bibinfo {author} {\bibfnamefont {H.}~\bibnamefont {Beck}},\ }\bibfield
  {title} {\bibinfo {title} {Fate of the excitonic insulator in the presence of
  phonons},\ }\href@noop {} {\bibfield  {journal} {\bibinfo  {journal}
  {Physical Review B}\ }\textbf {\bibinfo {volume} {90}},\ \bibinfo {pages}
  {195118} (\bibinfo {year} {2014})}\BibitemShut {NoStop}%
\bibitem [{\citenamefont {Kim}\ \emph {et~al.}(2021)\citenamefont {Kim},
  \citenamefont {Kim}, \citenamefont {Kim}, \citenamefont {Kwon}, \citenamefont
  {Kim},\ and\ \citenamefont {Kim}}]{kim2021direct}%
  \BibitemOpen
  \bibfield  {author} {\bibinfo {author} {\bibfnamefont {K.}~\bibnamefont
  {Kim}}, \bibinfo {author} {\bibfnamefont {H.}~\bibnamefont {Kim}}, \bibinfo
  {author} {\bibfnamefont {J.}~\bibnamefont {Kim}}, \bibinfo {author}
  {\bibfnamefont {C.}~\bibnamefont {Kwon}}, \bibinfo {author} {\bibfnamefont
  {J.~S.}\ \bibnamefont {Kim}},\ and\ \bibinfo {author} {\bibfnamefont
  {B.}~\bibnamefont {Kim}},\ }\bibfield  {title} {\bibinfo {title} {Direct
  observation of excitonic instability in ta2nise5},\ }\href@noop {} {\bibfield
   {journal} {\bibinfo  {journal} {Nature communications}\ }\textbf {\bibinfo
  {volume} {12}},\ \bibinfo {pages} {1969} (\bibinfo {year}
  {2021})}\BibitemShut {NoStop}%
\bibitem [{\citenamefont {Mazza}\ \emph {et~al.}(2020)\citenamefont {Mazza},
  \citenamefont {R{\"o}sner}, \citenamefont {Windg{\"a}tter}, \citenamefont
  {Latini}, \citenamefont {H{\"u}bener}, \citenamefont {Millis}, \citenamefont
  {Rubio},\ and\ \citenamefont {Georges}}]{mazza2020nature}%
  \BibitemOpen
  \bibfield  {author} {\bibinfo {author} {\bibfnamefont {G.}~\bibnamefont
  {Mazza}}, \bibinfo {author} {\bibfnamefont {M.}~\bibnamefont {R{\"o}sner}},
  \bibinfo {author} {\bibfnamefont {L.}~\bibnamefont {Windg{\"a}tter}},
  \bibinfo {author} {\bibfnamefont {S.}~\bibnamefont {Latini}}, \bibinfo
  {author} {\bibfnamefont {H.}~\bibnamefont {H{\"u}bener}}, \bibinfo {author}
  {\bibfnamefont {A.~J.}\ \bibnamefont {Millis}}, \bibinfo {author}
  {\bibfnamefont {A.}~\bibnamefont {Rubio}},\ and\ \bibinfo {author}
  {\bibfnamefont {A.}~\bibnamefont {Georges}},\ }\bibfield  {title} {\bibinfo
  {title} {Nature of symmetry breaking at the excitonic insulator transition:
  Ta 2 nise 5},\ }\href@noop {} {\bibfield  {journal} {\bibinfo  {journal}
  {Physical review letters}\ }\textbf {\bibinfo {volume} {124}},\ \bibinfo
  {pages} {197601} (\bibinfo {year} {2020})}\BibitemShut {NoStop}%
\bibitem [{\citenamefont {Windg{\"a}tter}\ \emph {et~al.}(2021)\citenamefont
  {Windg{\"a}tter}, \citenamefont {R{\"o}sner}, \citenamefont {Mazza},
  \citenamefont {H{\"u}bener}, \citenamefont {Georges}, \citenamefont {Millis},
  \citenamefont {Latini},\ and\ \citenamefont {Rubio}}]{windgatter2021common}%
  \BibitemOpen
  \bibfield  {author} {\bibinfo {author} {\bibfnamefont {L.}~\bibnamefont
  {Windg{\"a}tter}}, \bibinfo {author} {\bibfnamefont {M.}~\bibnamefont
  {R{\"o}sner}}, \bibinfo {author} {\bibfnamefont {G.}~\bibnamefont {Mazza}},
  \bibinfo {author} {\bibfnamefont {H.}~\bibnamefont {H{\"u}bener}}, \bibinfo
  {author} {\bibfnamefont {A.}~\bibnamefont {Georges}}, \bibinfo {author}
  {\bibfnamefont {A.~J.}\ \bibnamefont {Millis}}, \bibinfo {author}
  {\bibfnamefont {S.}~\bibnamefont {Latini}},\ and\ \bibinfo {author}
  {\bibfnamefont {A.}~\bibnamefont {Rubio}},\ }\bibfield  {title} {\bibinfo
  {title} {Common microscopic origin of the phase transitions in ta2nis5 and
  the excitonic insulator candidate ta2nise5},\ }\href@noop {} {\bibfield
  {journal} {\bibinfo  {journal} {npj Computational Materials}\ }\textbf
  {\bibinfo {volume} {7}},\ \bibinfo {pages} {210} (\bibinfo {year}
  {2021})}\BibitemShut {NoStop}%
\bibitem [{\citenamefont {Bruus}\ and\ \citenamefont
  {Flensberg}(2004)}]{bruus2004many}%
  \BibitemOpen
  \bibfield  {author} {\bibinfo {author} {\bibfnamefont {H.}~\bibnamefont
  {Bruus}}\ and\ \bibinfo {author} {\bibfnamefont {K.}~\bibnamefont
  {Flensberg}},\ }\href@noop {} {\emph {\bibinfo {title} {Many-body quantum
  theory in condensed matter physics: an introduction}}}\ (\bibinfo
  {publisher} {OUP Oxford},\ \bibinfo {year} {2004})\BibitemShut {NoStop}%
\bibitem [{\citenamefont {Coleman}(2015)}]{coleman2015introduction}%
  \BibitemOpen
  \bibfield  {author} {\bibinfo {author} {\bibfnamefont {P.}~\bibnamefont
  {Coleman}},\ }\href@noop {} {\emph {\bibinfo {title} {Introduction to
  many-body physics}}}\ (\bibinfo  {publisher} {Cambridge University Press},\
  \bibinfo {year} {2015})\BibitemShut {NoStop}%
\end{thebibliography}
%

	\appendix 
	\onecolumngrid
	\section{Detailed Calculation of the Photon Green's Function} \label{app.photongf}
	\begin{figure*}[h]
		\includegraphics[width=0.7\textwidth]{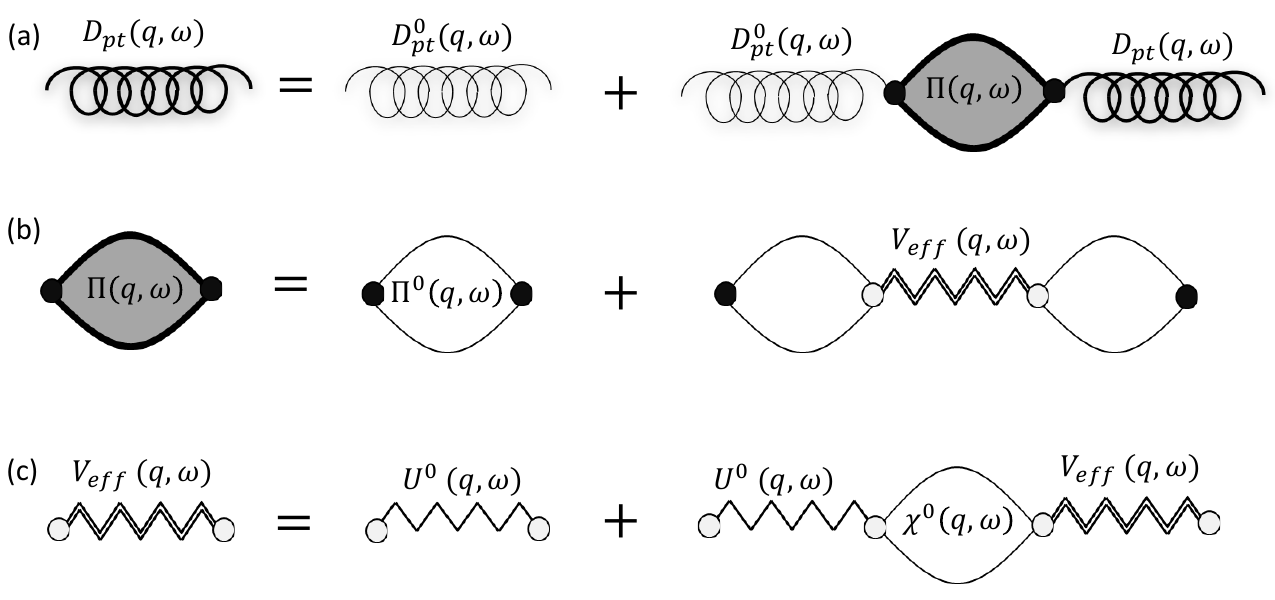}
		\caption{(a) Dyson equation for the photon Green’s function. (b) Equation for the photon self-energy \(\Pi(q,\omega)\). (c) Screened coulomb interaction, where \(\chi^0(q,\omega)\) represents the electronic polarization of the matter.} 
		\label{fig:phgf}
	\end{figure*}
	In this section, we will calculate the cavity photon Green's function, which is renormalized by the bare electron-photon and screened electron-electron interactions. The diagrammatic form of the Green's function is also shown in  Fig.~\eqref{fig:phgf}. Additionally, for the photon Green's function, we have:
	\begin{align}
		\mathcal{D}(q,\tau) = -\langle \mathrm{T}_\tau \hat{A}_q(\tau) \hat{A}_{-q}(0) \rangle, \label{eq:A1}
	\end{align}
	where \(\hat{A}_q(\tau) = \hat{a}^\dagger_q(\tau) + \hat{a}_q(\tau)\), \(\tau\) is the imaginary time, and \(\mathrm{T}_\tau\) is the imaginary time ordering operator. According to standard theoretical calculations, Eq. \eqref{eq:A1} can be solved as follows \cite{bruus2004many,coleman2015introduction}:
	\begin{align}
		\mathcal{D}(q,\tau)=-\frac{\sum_{n=0}^\infty\frac{(-1)^n}{n!} \int_0^\beta d\tau_1...\int_0^\beta d\tau_n\langle\mathrm{T}_\tau\hat{H}^{int}_{LM}(\tau_1)...\hat{H}^{int}_{LM}(\tau_n)\hat{A}_q(\tau)\hat{A}_{-q}(0)\rangle}{\sum_{n=0}^\infty \frac{(-1)^n}{n!}\int_0^\beta d\tau_1...\int_0^\beta d\tau_n\langle\mathrm{T}_\tau\hat{H}^{int}_{LM}(\tau_1)...\hat{H}^{int}_{LM}(\tau_n)\rangle}, \label{eq:A2}
	\end{align}
	The zeroth term of the above equation (n=0) is the bare photon Green's function
	\(\mathcal{D}_0(q,\tau)\), where in Matsubara frequency space
	\(\mathcal{D}_0(q,iq_m) = \int_0^\beta d\tau e^{iq_m\tau} \mathcal{D}_0(q,\tau)\), and by considering analytical continuation \(iq_m \rightarrow \omega + i0^+\) we have:
	\begin{align}
		\mathcal{D}_0(q,\omega) = \frac{2\omega(q)}{(\omega + i0^+)^2 - \omega(q)^2},
	\end{align}
	Here, 
	\(q_m = 2\pi m/\beta\), \(\beta\) is the inverse of the temperature, and \(m\) is an integer number.
	\(\omega(q)\) is the cavity photon energy before mixing with the state of the matter. The higher terms of Eq.\eqref{eq:A2} can be described by the second part of Fig.~\eqref{fig:phgf} (a), which shows the renormalization of the photon Green's function due to the presence of interactions in the system, that are contained in the photon self-energy \(\Pi(q,\omega)\). Thus, the full photon Green's function can be written as:
	\begin{align}
		\mathcal{D}(q,\omega) = \left[1 - \mathcal{D}_0(q,\omega) \Pi(q,\omega)\right]^{-1} \mathcal{D}_0(q,\omega),
	\end{align}
	According to Fig.\eqref{fig:phgf} (b), the photon self-energy contains two parts: \(\Pi(q,\omega)=\Pi_0(q,\omega)+\Pi_1(q,\omega)\).
	In addition, by considering the light-matter interaction Hamiltonian \eqref{eq:17}, in these diagram the black dots are showing the light-matter interaction strength which is given by \(\mathcal{G}_\nu(k,q)\)
	in the main text. So, the term \(\Pi^0(q,\omega)\) can be calculated as follows:
	\begin{align}
		\Pi^0(q,\omega)=\frac{1}{\beta}\sum_{\mu,\nu}\sum_k\sum_{\omega^\prime}\mathcal{G}_\mu(k,q)\mathcal{G}_\nu(k+q,-q)\mathrm{Tr}\left[\check{G}^0(k,\omega^\prime)\sigma_\mu\check{G}^0(k+q,\omega^\prime+\omega)\sigma_\nu \right],
	\end{align}
	where \(\check{G}^0(k,\omega) = \frac{1}{\omega - \hat{H}^{MF}_M(k) + i0^+}\) is the bare electronic Green's function. The second term of the photon self-energy contains the renormalization due to the screened electron-electron interaction, shown by the vertex correction \(\check{V}_{eff}(q,\omega)\), which can be calculated using the RPA approach. For the vertex correction term if we rewrite the electron coulomb interaction term \eqref{eq:4} in the basis of the density operator \(\hat{\rho}_{k,\nu}(q)=\Psi_k^\dagger \hat{\sigma}_\nu \Psi_{k+q}\), we have \(\hat{H}_{int}=\sum_{k,q}\sum_{\nu,\mu}\hat{\rho}_{k,\mu}(q) \check{U}^0\hat{\rho}_{k,\nu}(-q)\), with  \(\check{U}^0 = \frac{V}{2}\mathrm{diag}(1, -1, -1, -1)\), so  according to Fig.\eqref{fig:phgf} (c), the effective interaction \(\check{V}^{\text{eff}}(q,\omega)\) will become as follows:
	\begin{align}
		\check{V}^{\text{eff}}(q,\omega) = \frac{\check{U}^0}{1 - \check{U}^0\chi^0(q,\omega)},
	\end{align}
	In equation above, \(\chi^0(q,\omega)\) is the 0th order RPA bubble diagram, and can be calculated as:
	\begin{align}
		\chi^0(q,\omega)=\frac{1}{2\pi}\int dk \sum_{\alpha,\beta}\frac{f(E^\alpha_k,T)-f(E^\beta_{k+q},T)}{E^\alpha_k-E^\beta_{k+q}+\omega+i0^+}\langle \alpha | \sigma_\mu | \beta \rangle\langle \beta | \sigma_\nu | \alpha \rangle,
	\end{align}
	Building on this, the second term of the photon self-energy \(\Pi^1(q,\omega)\) can be derived as:
	\begin{align}
		&\Pi^1(q,\omega)=\frac{1}{\beta^2}\sum_{\mu,\nu}\sum_{k,k^\prime}\sum_{\omega^\prime,\omega^{\prime\prime}}\sum_{\mu^\prime,\nu^\prime} \mathcal{G}_\nu(k+q,-q)\mathrm{Tr}\left[\check{G}^0(k^\prime,\omega^{\prime\prime})\sigma_{\nu^\prime} \check{G}^0(k^\prime+q,\omega^{\prime\prime}+\omega)\sigma_\nu\right]\nonumber\\
		&~~~~~~~~~~~~~~~~~~~~~~~~~~~~~~~~~~~~~~~~~~~~~~~\times \check{V}_{\nu^\prime\mu^\prime}^{eff}(q,\omega)\mathrm{Tr}\left[\check{G}^0(k,\omega^\prime)\sigma_\mu\check{G}^0(k+q,\omega^\prime+\omega)\sigma_{\mu^\prime}\right]\mathcal{G}_\mu(k,q).
	\end{align}
	
\section{Cavity Photon Spectral Function: Analysis of \(\mathrm{Im}\left[\mathcal{D}(q,\omega)\right]\)}\label{app.imagGF}
\begin{figure*}[t]
		\includegraphics[width=0.9\textwidth]{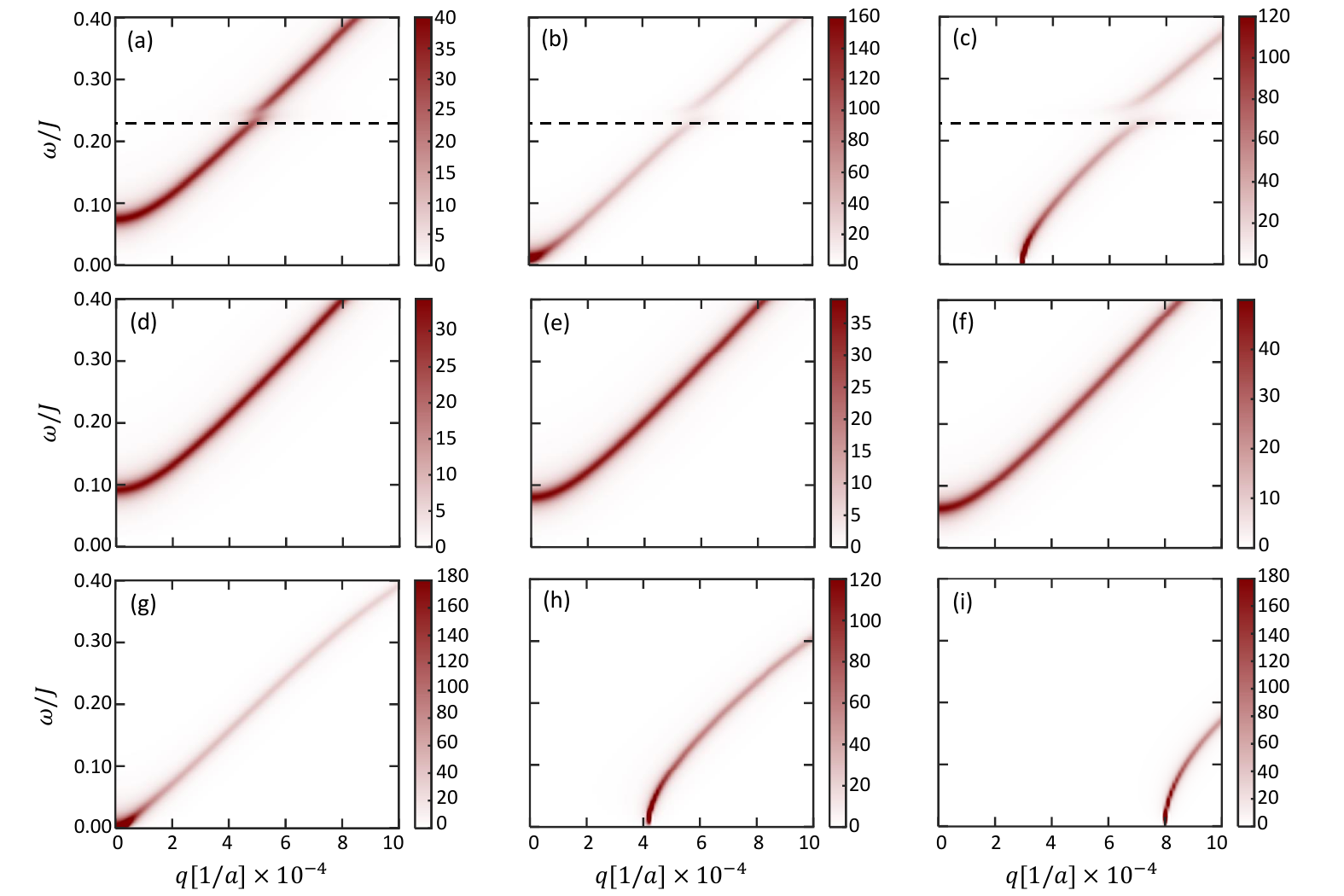}
		\caption{Spectral densities  \(-\Im[\mathcal{D}(q,\omega)]\) for various cases are presented to complement Fig.~\ref{fig:Dpt} in the main text. We adopt natural units with \(\hbar=6.6\times10^{-16} \mathrm{eV.s}\) and \(c=3\times10^8 \mathrm{ m/s}\) and set \(\hbar\omega_c/J=0.1\); other parameters are \(J=0.1\mathrm{eV}\) and \(a=4 \text{\AA}\). Rows correspond to the excitonic, trivial, and topological insulating phases; columns increase in light–matter coupling \(g\) from 0.2 to 0.4. In panels (a–c), the dashed black line marks the excitonic phase mode energy, approximately \(\omega_{\mathrm{phase}}/J=0.23\).} 
		\label{fig:ImDpt}
	\end{figure*}

In general, the identification of energy poles is most directly obtained by examining the imaginary part of the Green’s function. In the main text, motivated by the heterodyne photodetection setup, we focused on the real part of the Green’s function, which reveals the energy dispersion of the cavity photon mode. Here, in this appendix, we complement that analysis by emphasizing the intrinsic connection between the real and imaginary parts of the Green’s function, governed by the Kramers–Kronig relation. According to this causality-based relation, the imaginary part corresponds to energy dissipation and is directly linked to the location of zero-crossings in \(\Re[\mathcal{D}(q,\omega)]\) across positive and negative frequencies, as illustrated in Fig.~\eqref{fig:ImDpt}.
	
\section{Excitonic Collective Modes}\label{app.collective mode}
		We probe the system’s collective excitations via the retarded density–density response function, defined by
		\begin{align}
			\chi_{\mu\nu}=-i\theta(t)\frac{\langle\psi_0|\left[\hat{\rho}_\mu(t),\hat{\rho}_\nu(0)\right]|\psi_0\rangle}{\langle\psi_0|\psi_0\rangle},
	\end{align}
	\begin{figure*}[t!]
		\includegraphics[width=0.7\textwidth]{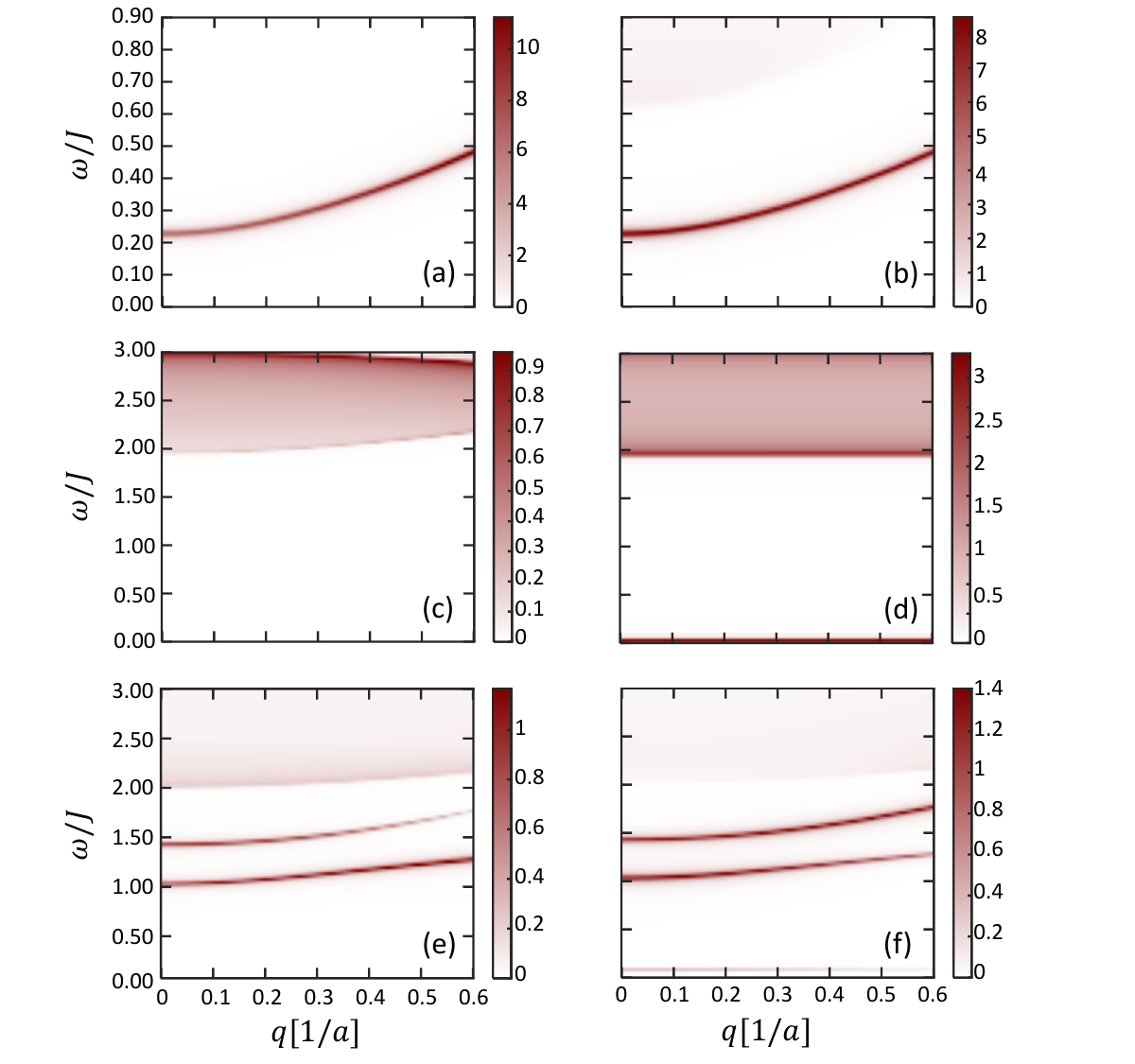}
		\caption{Collective mode spectrum of the excitonic insulator over a large range of momenta: each row corresponds to one parameter set—panels (a,b) to the purely electronic case with \( V/J=2.5\), \(D/J=1.1\), \(J_{sp}/J=0.5\), \(\phi=0.116\); (c,d) to the primarily lattice case with \( V/J=0.1\), \(D/J=0.5\), \(J_{sp}/J=0.5\), \(\lambda/J=1.4\), \(\phi=0.075\); and (e,f) to the primarily electronic case with \( V/J=2.8\), \(D/J=0.5\), \(J_{sp}/J=0.5\), \(\lambda/J=0.1\), \(\phi=0.1755\). Panels (a,c,e) display \(\-\mathrm{Im}[\chi_{22}]/\pi\) (phase mode), while (b,d,f) display \(\-\mathrm{Im}[\chi_{11}]/\pi\) (amplitude mode). In the primarily lattice and primarily electronic cases the single-particle gap \(E_g/J\) is fixed at 2 (\(J = 0.1 \mathrm{eV}\)).}\label{fig:collective mode}
\end{figure*}
where \(\theta(t)\) is the Heaviside step function, \(|\psi_0\rangle\) denotes the interacting ground state, and \(\hat{\rho}_\mu(t)\)  is the density operator in the Heisenberg picture. 
		Fourier transforming the above expression to momentum and frequency space and invoking the RPA yields
		\begin{align}
			\chi(q,\omega)=\frac{\chi^0(q,\omega)}{1-\chi^0(q,\omega)\check{U}^0}
		\end{align}
		In this expression,  \(\check{U}^0 = \mathrm{diag}(\frac{V}{2}, -\frac{V}{2}+\lambda^2 \mathcal{D}^0_{\text{pn}}(q,\omega), -\frac{V}{2}, -\frac{V}{2})\) defines the effective electron–electron interaction, combining the bare Coulomb interaction with the phonon-mediated electron interaction.
	
The resulting excitonic collective modes —introduced in Sections \ref{sec:hybrid_modes} and \ref{sec:phonon} of the main text— are plotted in Fig. \eqref{fig:collective mode} via the spectral functions  (\(-\mathrm{Im}[\chi_{11}]/\pi\)) for the amplitude channel, and  (\(-\mathrm{Im}[\chi_{22}]/\pi\)) for the phase channel, using a broadening factor \(\eta=0.01\). The well-defined low-energy dispersions in Fig. \eqref{fig:collective mode}(a,b) and Fig. \eqref{fig:collective mode}(e,f) result from the collective excitonic modes. However, the upper dispersion in \eqref{fig:collective mode}(e,f) is due to the interband transtions with BCS-type dispersion as explained in \cite{khatibi2020excitonic}.
	
\end{document}